\documentclass[twocolumn]{aastex62}
\usepackage{amsmath}
\graphicspath{{./}{figures/}}

\received{2019 July 7}
\revised{2019 September 17}
\accepted{2019 September 18}
\submitjournal{ApJ}

\shorttitle{The Nature of Class I Sources}
\shortauthors{Ellithorpe et al.}
\begin{document}

\title{The Nature of Class I Sources: Periodic Variables in Orion}

\correspondingauthor{Elizabeth A. Ellithorpe}
\email{ellithorpe@ou.edu}

\author{Elizabeth A. Ellithorpe}
\affiliation{Physics and Astronomy Department, University of Oklahoma, Norman, OK 73019 USA}
\affiliation{Astronomy Department, University of California, Berkeley, CA 94720, USA}

\author{Gaspard Duchene}
\affiliation{Astronomy Department, University of California, Berkeley, CA 94720, USA}
\affiliation{Universit\'e Grenoble-Alpes, CNRS Institut de Plan\'etologie et d'Astrophysique (IPAG), F-38000 Grenoble, France}

\author{Steven W. Stahler}
\affiliation{Astronomy Department, University of California, Berkeley, CA 94720, USA}

\nocollaboration



\begin{abstract}

We present a quantitative, empirically based argument that at least some Class~I sources are low-mass, pre-main-sequence stars surrounded by spatially extended envelopes of dusty gas. 
The source luminosity arises principally from stellar gravitational contraction, as in optically visible pre-main-sequence stars that lack such envelopes. We base our argument on the fact that some Class~I sources in Orion and other star-forming regions have been observed by {\it Spitzer} to be periodic variables in the mid-infrared, and with periods consistent with T~Tauri rotation rates. Using a radiative transfer code, we construct a variety of dust envelopes surrounding rotating, spotted stars, to see if an envelope that produces a Class~I SED 
at least broadly matches the observed modulations in luminosity.
Acceptable envelopes can either be spherical or flattened, and may or may not have polar cavities. The key requirement is that they have a modest equatorial optical depth at the {\it Spitzer} waveband of 3.6~$\mu$m, typically $\tau _{3.6} \approx 0.6$. 
The total envelope mass, based on this limited study, is at most about 0.1~$M_\odot$, less than a typical stellar mass. 
Future studies should focus on the dynamics of the envelope, to determine whether material is actually falling onto the circumstellar disk.

\end{abstract}



\section{Historical Introduction}

The evolutionary status of young stars embedded in dusty gas remains problematic. Traditionally, researchers have utilized the observed spectral energy distribution (SED) of a source to infer the age of its central star \citep[e.g.,][]{eva09}. The youngest sources, dubbed Class~0, lack both optical and near-infrared flux, and their SEDs continue to rise beyond 100~$\mu$m \citep{and93}. More quantitatively, Class~0 sources have the lowest bolometric temperature, $T_{\rm bol}$, where $T_{\rm bol}$ is a global measure of the envelope's mean dust temperature \citep{mye93}. They also have submillimeter luminosities (\hbox{$\lambda\,>\,350\,\,\mu{\rm m}$}) greater than about 0.5~percent of the bolometric value, indicating that the mass of the star's envelope exceeds that of the star itself \citep{and93}.

Within this same empirical framework, the next youngest category is Class~I. Here, there is detectable near-infrared flux, and the SED rises from 2.2 to 20~$\mu$m. The envelope appears to contain less mass than the star, since the millimeter luminosity is a smaller fraction of $L_{\rm bol}$ than in Class~0 sources. By now, it is common practice for observers to refer to both Class~0 and Class~I objects as ``protostars" \citep[e.g.,][]{fur16}, and to assume that the latter are, in some sense, more evolved than the former. 

Exactly what this sense is remains unclear. 
Over the past 15~years, many observers have probed various features of Class~I sources, in part to address this evolutionary
issue. \cite{whi04} observed 15 Class~I stars in Taurus optically, via scattered light. Spectroscopy indicated that both the stars' effective temperature and rotation speeds were indistinguishable from the 
optically revealed Class~II sources, long identified as more evolved, pre-main-sequence stars. 
However, the target sources could have been a relatively old subset of Class~I sources, since they had envelopes that were porous enough to allow the escape of optical photons. Using a broader sample, and focusing on the near-infrared spectrum, \cite{cov05} found a larger mean rotation speed than in Class~II objects. On the other hand, \cite{dop05} found the stellar luminosity of these objects to be very similar to classical T~Tauri stars. It should be noted that obtaining $L_\ast$ in any source pivots on a derived value for the extinction. Unfortunately, obtaining this extinction is highly problematic in Class~I, a point emphasized by \cite{bec07}.

\cite{pra09} and \cite{con10} estimated Class~I mass accretion rates, using the observed flux in the near-infrared Br$\gamma$ line and an empirical relationship initially established for T~Tauri stars \citep{muz98}. Most of the resulting values for the accretion luminosity $L_{\rm acc}$ were far less than the total luminosity of the source. But even this point remains controversial. To cite one specific example, \cite{bec07} used Br$\gamma$ emission to estimate $L_{\rm acc} / L_{\rm bol} = 0.008$ for the source IRAS~04016+2610 in Taurus. \cite{fur08}, who fit a multi-parameter model to the star's SED, derived $L_{\rm acc} / L_{\rm bol}\,=\,0.90$.

Modeling the SED yields, at least potentially, both the mass of the envelope and the rate at which gas is falling onto the star. Matching the observed SEDs with those from models of collapsing, rotating clouds suggests Class~I infall rates of order \hbox{$10^{-6}\,\,M_\odot\,\,{\rm yr}^{-1}$} (\cite{whi03}, two orders of magnitude higher than those typically obtained for Class~II objects \citep{har16}. On the other hand, the infall rates derived via infrared spectroscopy are lower by several orders of magnitude  \citep[e.g.,][]{pra09}. There is still no generally accepted resolution to this discrepancy.

Fortunately, the {\it temporal} behavior of Class\,I sources provides an important clue to the nature of the central object, and therefore to the evolutionary state. Low-mass, pre-main-sequence stars that do {\it not} reside in dusty envelopes are well known for their variability, which can be either erratic or quasi-periodic \citep[e.g.,][]{cod14}. In the latter case, hot or cool spots on the stellar surface, rotating in and out of view, plausibly explain the luminosity fluctuations, although asymmetric disk emission can also contribute at infrared wavelengths. 

In recent years, {\it Spitzer} monitoring campaigns of star-forming regions have revealed that some Class\,I sources also display quasi-periodic photometric variations \citep[e.g.,][]{mor11,wol15}, further suggesting that these objects could be spotted pre-main sequence stars. However, the interpretation of this finding is not straightforward, as the infrared light is reprocessed by the circumstellar envelope. Thus, one might expect that the spot emission would be redistributed in a more isotropic fashion and toward longer wavelengths, suppressing the characteristic near-infrared photometric signal. This effect strongly depends on the envelope structure and total optical depth. Therefore, to properly interpret the implications of the periodic variability of Class\,I sources, it is necessary to assess how deeply embedded they truly are.

The present paper develops this line of argument quantitatively. We employ a radiative transfer code to explore numerical models of dust envelopes surrounding rotating, spotted stars. We test these models against observational data, consisting of a set of periodically varying Class~I sources in Orion. We judge models to be acceptable only if they {\it both} yield Class~I SEDs {\it and} infrared light curves that resemble those of our observed sample.  This double constraint allows us to characterize important features of the envelopes' density distribution in more detail than previous studies. 

Our main result is that Class\,I envelopes, while sufficiently opaque to block starlight from reaching the observer directly, must have a modest mid-infrared optical thickness \hbox{($\tau_{3.6} \,\approx\, 0.6$)} to preserve an intrinsic periodic signal. 
  
Flattened envelopes similar to those in rotating, infalling clouds are acceptable, if the envelope mass is sufficiently low. So are purely spherical clouds with power-law density profiles, as empirically found in Class\,0 envelopes \citep[e.g.,][]{tob15}, again provided that the mid-infrared optical thickness is modest. Neither SED analysis nor photometric lightcurves allows us to constrain the detailed struture of the envelope, which can only be determined via future high-resolution imaging studies.

In any event, since the observed periodicity of the Orion Class\,I sources is compatible with stellar rotation rates, our findings add weight to the possibility that Class~I sources generally are embedded pre-main-sequence stars. 

We begin, in Section\,\ref{sec:env_models} below, by exploring which dust envelopes yield appropriate Class~I SEDs. From a purely practical standpoint, we require such an envelope for our analysis of the periodic behavior of an embedded, spotted Class~I object. To elucidate the best-fitting model envelope, we use a radiative transfer code to produce synthetic SEDs, which we compare to an empirical, Class~I template. In Section\,\ref{sec:ClassI}, we review the {\it Spitzer} observations of periodically varying Class~I sources in Orion. We display the SEDs and construct light curves for all stars. We account for these light curves by assuming there are single spots located on the rotating stellar surface. We then utilize our previously constructed set of envelopes yielding proper SEDs to see which models also produce satisfactory light curves. We also review, for comparison, the observations of periodic Class II and III sources; our Orion Class~I sample is more akin to the former than to the latter. Section\,\ref{sec:discus} summarizes our findings and discusses the broader astrophysical implications of this work.

\section{Class I Envelopes}
\label{sec:env_models}

\subsection{Modeling strategy}

\begin{figure}
\plotone{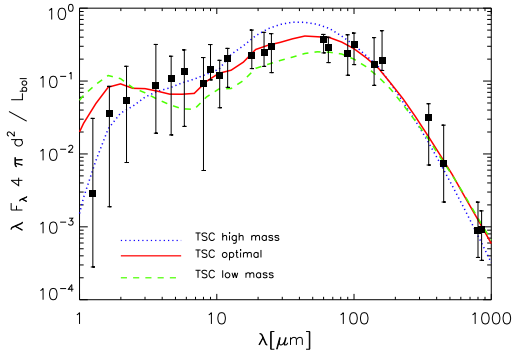}
\caption{Luminosity-normalized SEDs generated by three TSC-type envelopes, along with the empirical Class~I template SED. Shown are SEDs in the best-fitting case, along with those from the lowest and highest-mass acceptable envelope masses. Parameters for all three models are listed in Table\,\ref{tab:models}.\label{fig:tsc_model}}
\end{figure}

\begin{figure}
    \plotone{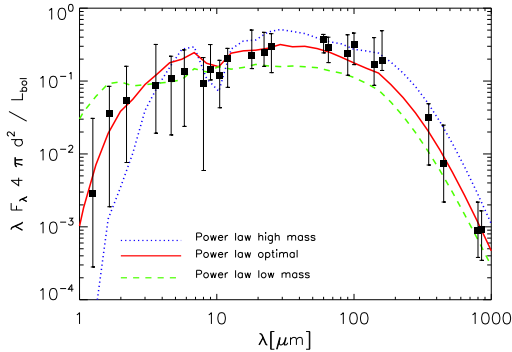}
    \caption{SEDs generated by three  power-law envelopes ($\gamma=1.0$, along with the template Class~I SED. Shown are SEDs from the optimal, lowest-mass, and highest-mass model envelopes.\label{fig:powerlaw_model}}
\end{figure}

We first briefly examine which dusty envelopes surrounding a young star produce an SED typical of Class~I sources. To this end, we compare synthetic SEDs to observed ones. Empirically, there is a rather broad range of SED morphologies that all share the defining Class~I property of a rising flux between 2 and 20~$\mu$m. We have therefore constructed a representative, or template, SED by suitably averaging photometric data from 21 sources in Taurus, the star-forming region with the best-studied population of Class\,I sources. To combine the SEDs, we first normalized each to the bolometric luminosity of the system, so as to focus on the intrinsic shape of the flux distribution. We refer the reader to Appendix\,\ref{appendix} for the remaining details of our procedure for constructing the template. We also built a template based on Orion Class\,I sources, which is similar to the Taurus one, albeit with some biases introduced by crowding and confusion. We therefore adopt the Taurus template, illustrated in Figures\,\ref{fig:tsc_model} and \ref{fig:powerlaw_model}, for our analysis.

For constructing models of Class\,I sources, we assume three components: a pre-main sequence star at the center, a circumstellar disk, and an extended envelope. The parameters describing each component are presented in the remainder of this section. Each synthetic SED was obtained using the 2D version of the radiative transfer code \hbox{MCFOST} \citep{pin06,pin09}. To evaluate whether a given model adequately matches the template Class\,I SED, we used a standard \hbox{$\chi^2$-test}. We account for the asymmetric errorbars associated with the template by adopting the upper or lower uncertainty depending on whether the model over- or under-predicts the flux of the template at a given wavelength. We consider models yielding reduced $\chi^2_{\rm red} \leq 1$ to be satisfactory matches to the template SED. (Of the objects used in defining the template, two thirds pass this criterion when compared to the template itself.) 

The parameter space describing a Class\,I system is impractically large to perform a thorough explorations. Besides the sheer number of parameters being a limiting factor in how detailed our analysis can be, SEDs are notoriously degenerate, making convergence on a single model difficult. We therefore decided to fix as many parameters as possible, and only allow the total disk and envelope masses to vary in our investigation of acceptable models of Class\,I systems. In all models, we assume that dust comprises 1~percent of the gas mass. To evaluate the SED of the system, we also need to select a viewing angle. We choose a polar angle of 
\hbox{$\theta\,=\,57^\circ$}, which is the mean value for an envelope with random orientation.

For the central star, we adopt a total luminosity of 0.9~$L_\odot$. We further assume that the spectrum of the object is well represented by a 4000\,K NextGen stellar atmospheric model \citep{allard12}, typical for Class\,II and III sources in star-forming regions.
Thus, the stellar radius is 2~$R_\odot$. For our radiative transfer calculation, we need not specify the stellar mass. However, we do require this quantity in one of our envelope models (see below), and for that purpose we choose the representative value
\hbox{$M_\ast\,=\,0.5\,M_\odot$}. 

It has long been known that, in regions as young as Orion, the vast majority of optically visible pre-main-sequence stars are surrounded by circumstellar disks \citep{h01}. We therefore assume that our more embedded stars also have such disks. Our model disks have a surface density that varies as 
$r^{-3/2}$, a 100:1 gas-to-dust mass ratio,
and a standard flared structure. Thus, the scale height $H$ is given by \hbox{$H_0 (r/r_0)^{6/5}$}, where $H_0=10\,{\rm au}$ and $r_0$ is the outer radius of $100\,{\rm au}$. The inner radius is 0.073~au, which is the representative value adopted by \cite{whi03} for the location where dust is thermally sublimated. The dust itself is characterized by a 3.5:1 silicate/graphite mixture and a power-law size distribution extending from 0.03 to 1~mm, with an index of -3.5 \citep{mrn}. Finally, we assume that the disk itself is passively irradiated by the star, i.e., we do not include any additional luminosity either from infall onto the disk surfaces or from accretion within that structure.

In exploring the envelope structure, we consider two model families (see below) as in some previous studies of Class\,I sources \citep[e.g.,][]{fur08}. Both families have the same inner and outer radii, 0.073 and 5000\,au, respectively. Here, we are guided by the SED study of \cite{whi03}. We further note that the outer radius has no significantly influence on the resulting SED, whereas the inner radius must not exceed $\sim1\,$au in order to match the Class\,I template SED in the 1--5\,$\mu$m range. The dust in our model envelopes has the same compositional mixture as for the disk, but a size distribution extending from 0.03 to 1\,$\mu$m with a -4.5 power law. We again assume a 100:1 gas-to-dust ratio. These parameters were chosen so as to produce an extinction law that matches observations in the interstellar medium and that was used in Class\,I models by \cite{whi03}\footnote{More recent observations indicate that the extinction curve in molecular clouds is flatter than that in the interstellar medium for wavelengths $\lambda \gtrsim 3\,\mu$m \citep{fla07}.}. 

\subsection{TSC envelopes}

For over 30~years, many researchers seeking to model non-spherical envelopes around young stars have utilized the classic study of \citet{cas81}. These authors calculated the density distribution in the deep interior of a collapsing, rotating cloud. The cloud itself is the highly idealized singular, isothermal sphere, whose self-similar collapse was previously studied by \citet{shu77}. \citet{cas81} found an analytic expression for the density as a function of radius $r$ and polar angle $\theta$: 
\begin{equation}
\begin{split}
& \rho(r,\theta)=\frac{\dot{M}_{\rm env}}
{4\pi(GM_{*}R^3_c)^{1/2}}\left(\frac{r}{R_c}\right)^{-3/2} 
\\ &
\times \left(1+\frac{\mu}{\mu_0}\right)^{-1/2}\,\left(\frac{\mu}{\mu_0}+\frac{2\mu  R_c}{r} \right)^{-1}. 
\end{split}
\end{equation}

In this equation, the quantity $R_c$ represents the centrifugal radius, i.e., the outermost point where fluid trajectories land in the equatorial plane. The density is peaked toward the central star, varying with radius as $r^{-1/2}$ for $r \ll R_c$ and as $r^{-3/2}$ for $r \gg R_c$. A second quantity appearing in the equation is \hbox{${\dot M}_{\rm env}$}, the total mass accretion rate onto the disk and star, whose mass is $M_\ast$. The expression also contains the variables
\hbox{$\mu\,\equiv\,{\rm cos}\,\theta$} and $\mu_0\,\equiv\,{\rm cos}
\,\theta_0$, where $\theta_0$ is the polar angle at which each fluid element leaves the rarefaction wave. Since the element subsequently traces a parabola, $\mu$ is a function both of $r$ and the parameters $\mu_0$ and $R_c$. This functional dependence is given implicitly by
\begin{equation}
\mu_0^3+\mu_0\left(\frac{r}{R_c}-1\right)-\mu\left(\frac{r}{R_c}\right)=0 .
\end{equation}

The studies by \citet{cas81} and \citet{shu77}, provide expressions for both ${\dot M}_{\rm env}$ and $R_c$. The first is a function only of the parent cloud temperature (assumed to be uniform), while the second also depends on the cloud's rotation rate and the time since the start of collapse. In practice, researchers have treated 
${\dot M}_{\rm env}$ and $R_c$ as free
parameters, thereby generating families of envelope solutions. These generalized models, which we designate as TSC, are {\it not} physical solutions to the problem of a rotating, collapsing cloud. They are simply a set of non-spherical envelopes useful for characterizing observed sources.

We follow this practice here, while noting, for example, that
${\dot M}_{\rm env}$ cannot be interpreted literally as the infall rate in the envelope; it is simply a scale factor in the expression for envelope density. In any event, we implemented equations~(1) and (2) to construct a dust envelope model for use in MCFOST. Following \cite{whi03}, we set \hbox{$R_c = 100~{\rm au}$}. With these values for $R_c$ and the inner and outer radii of the envelope, the total mass of the envelope is proportional to the mass accretion rate in the envelope, with 
\begin{equation}
M_{\rm env} = 0.19\ \left(\frac{{\dot M}_{\rm env}}{5\times 10^{-6}\,M_\odot.{\rm yr^{-1}}}\right)\, M_\odot .
\end{equation}

\begin{table}
\centering
\caption{Parameters of synthetic models that assume a TSC envelope (top part of the table) and a power law envelope (PL, with $\gamma=1$, bottom part of the table). The total gas mass in the envelope and disk are listed in columns 2 and 3, respectively. Column 4 indicates the 3.6\,$\mu$m optical depth through the envelope; for the TSC envelope, the lowest (highest) optical depth is found at the pole (equator). Column 5 presents the $\chi^2_{\rm red}$ when compared to the Class\,I template.}
\label{tab:models}
\begin{tabular}{lcccc}
\hline
\hline
Model & $M_{\rm env}\ (M_\odot)$ & $M_{\rm disk}\ (M_\odot)$ & $\tau_{3.6}^{\rm env}$ & $\chi^2_{\rm red}$ \\
\hline
TSC low-mass & 2.7\,10$^{-2}$ & 3\,10$^{-3}$ & 0.3--1.0 & 0.95 \\
TSC optimal & 5\,10$^{-2}$ & 2\,10$^{-3}$ & 0.6--1.9 & 0.23 \\
TSC high-mass & 1\,10$^{-1}$ & 5\,10$^{-4}$ & 1.1--3.8 & 1.04 \\
\hline
PL low-mass & 5\,10$^{-2}$ & 3\,10$^{-3}$ & 0.6 & 0.94 \\
PL optimal & 1.25\,10$^{-1}$ & 1\,10$^{-3}$ & 1.5 & 0.14 \\
PL high-mass & 2.75\,10$^{-1}$ & 3\,10$^{-5}$ & 3.2 & 0.98 \\
\hline 
\end{tabular}
\end{table}

The solid curve in Figure~1 displays the best-fit SED, along with the discrete points and error bars corresponding to our Class~I template. 
For this model, the envelope mass was
\hbox{$M_{\rm env}\,=\,0.050\,M_\odot$}, and the disk mass was \hbox{$M_d \,=\, 2\times 10^{-3}\,\,M_\odot$}. Note that \hbox{$M_{\rm env}\,<\,M_\ast$},
a property that we found to hold generally, in line with previous studies of the SED of Class\,I sources \citep[e.g.,][]{whi03, fur16}. The optimal TSC model has $\chi^2_{\rm red}=0.23$. 

A key property of the optimal model is its modest optical depth at mid-infrared wavelengths. Referring to the {\it Spitzer} waveband of 3.6\,$\mu$m, we find $\tau_{3.6}\,=\,0.8$ at our viewing angle of $57^\circ$. The corresponding figure is 0.6 (1.9) along the pole (equator).  
As we will see, limited optical depth turns out to be the essential property for reproducing a Class~I-type SED. 

To illustrate this point further, we also display in Figure~1 models that have the lowest and highest acceptable envelope mass; these both have $\chi^2_{\rm red}$-values close to unity. The dashed SED \hbox{($\chi^2_{\rm red}\,=\,0.95$)} was obtained from a model with
\hbox{$M_{\rm env}\,=\,0.027~M_\odot$} and
\hbox{$M_d\,=\,3\times 10^{-3}~M_\odot$}. Here,
\hbox{$\tau_{3.6}\,=\,0.4$}
at \hbox{$\theta\,=\,57^\circ$}. 
With such a relatively low optical depth, the short-wavelength radiation from the star and inner disk partially penetrates the envelope, lifting the SED above the template in this regime. If we were to include an outflow cavity, as was done by \citet{whi03} in several models, then the short-wavelength SED would be even more elevated.
 
Finally, the dotted curve in Figure~1 corresponds to our highest acceptable envelope mass,
\hbox{$M_{\rm env}\,=\,0.1~M_\odot$}. In this case,
the SED has \hbox{$\chi^2_{\rm red}\,=\,1.04$}. The
disk mass is actually lower than for the optimal model,
\hbox{}{}{$M_d\,=\,5\times 10^{-4}~M_\odot$}. The short-wavelength portion of the SED mimics the template quite well, but rises above it at longer wavelengths, representing absorption and reemission of radiation by the relatively optically thick envelope.

We see that dust envelopes constructed according to the TSC model yield SEDs whose {\it shapes} can match typical Class~I sources. However, the {\it luminosities} predicted by these same models are too large, by an order of magnitude. Thus, the accretion luminosity corresponding to our optimal TSC model, \hbox{$L_{\rm acc}\,=\,G M_\ast {\dot M_{\rm env}} / R_\ast$} is $10\,L_\odot$,
{\it provided} we interpret the parameter
${\dot M}_{\rm env}$ as an infall rate.
In contrast, the median bolometric luminosity of the Taurus Class\,I sources used in generating the template is only $0.8~L_\odot$. 
This discrepancy is a manifestation of the venerable ``luminosity problem"
\citep{kh95}. However, the paradox disappears if we recognize, as already stressed, that ${\dot M}_{\rm env}$ in TSC models is simply a parameter governing the envelope mass, and does not reflect the true infall rate, if any.

\subsection{Power-law envelopes}

TSC models resemble, at least qualitatively, rotating clouds undergoing inside-out collapse. However, the inferred envelope masses that yield Class~I SEDs are much smaller than observed values for dense cores. We noted that a TSC envelope yields an acceptable SED only when the mid-infrared optical thickness is modest. It is {\it this} feature, rather than the dynamics of the original TSC theory, which is essential.

We bolster this statement by considering envelopes that, unlike TSC models, have no polar variation in density. That is, we assume the density to be spherically symmetric, and to vary as a power law in radius: \hbox{$\rho\,\propto\,r^{-\gamma}$}. In addition to their simplicity, such models have some empirical motivation. %As mentioned earlier, 
Resolved millimeter maps of Class~0 sources find density variations that are at least consistent with power-laws. Furthermore, there is no indication of a change in the index at an $R_c$-value of order 100~au \citep[e.g.,][]{clt12,tob15,per16}. Unfortunately, such finer mapping is not yet feasible for the lower-mass Class~I envelopes, so the finding for less-evolved objects is only suggestive. 

Keeping the previous star and disk models, we now construct power-law envelopes that still have inner and outer radii of 0.07~au and 5000~au, respectively. For the exponent $\gamma$, we choose 1. This value is broadly consistent with the millimeter observations
\citep[e.g.,][]{tob15,per16}, and differs from the $\gamma$ of 1/2 that characterizes the deep interior of TSC models. 

We display the SED generated by the optimal power-law envelope as the solid curve in Figure~2. It is evident visually that the match to the template SED is quite close; the associated $\chi_{\rm red}^2$-value is 0.14, even smaller than for the optimal TSC model. In this case, $\tau_{3.6}\,=\,1.5$ at all polar angles. The mass of the envelope is now 0.13~$M_\odot$, while that of the disk is $1\times10^{-3}~M_\odot$. 

The dashed curve in Figure~2 shows the SED for the lowest-mass envelope that still yields an acceptable match to the Class~I template, i.e., with a $\chi^2_{\rm red}$ near unity. Here, \hbox{$M_{\rm env}\,=\,0.05~M_\odot$}, \hbox{$M_d\,=\,1\times 10^{-3}~M_\odot$}, and \hbox{$\chi^2_{\rm red}\,=\,0.94$}. 
The optical thickness $\tau_{3.6}$ is 0.6. As in the analogous TSC model, the short-wavelength portion of the synthetic SED rises markedly above the template. In the maximum-mass envelope, whose SED is shown as the dotted curve, the short-wavelength portion is suppressed, while the SED rises in the mid-infrared region with a marked silicate absorption feature which is present in only some Class\,I sources \citep{fur08}. Here, \hbox{$M_{\rm env}\,=\,
0.28~M_\odot$}, \hbox{$M_d\,=\,3\times 10^{-3}~M_\odot$}, and 
$\chi^2_{\rm red}\,=\,0.98$. The relevant optical depth is now \hbox{$\tau_{3.6}\,=\,3.2$}.

To compare our model envelope masses, both TSC and power-law, with observed ones, we note that the $K$-band extinction is a convenient proxy for the latter. The dust extinction law adopted for the envelope is such that $A_K / \tau^{\rm env}_{3.6} \approx 3.3$. Thus, the typical line-of-sight extinction associated with Class\,I envelopes is a few magnitudes at $K$. This is fully consistent with the empirical estimates based on spectral and photometric analysis of known Class\,I systems \citep{pra09, con10, bec07, fur16}.

\section{Periodic Class~I sources in Orion}
\label{sec:ClassI}

\subsection{Source list}

\begin{figure}
    \plotone{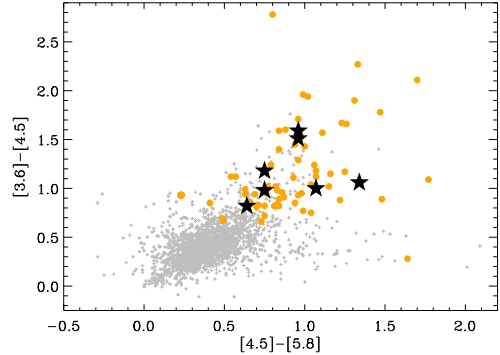}
    \caption{Color-color diagram of young stellar objects in Orion based on {\it Spitzer}/IRAC photometry. Gray plus signs and orange circles represent Class\,II sources and Class\,I sources,
    respectively, as identified by \citep{meg12}. Large black stars represent the sources considered in this study. \label{fig:spitzer_ccd}}
\end{figure}

This study was initially motivated by the identification of several periodically variable Class I sources in Orion, based on a 40\,d 3.6 and 4.5\,$\mu$m photometric monitoring campaign with {\it Spitzer} \citep{mor11}. To confirm the objects' class and status as periodic variables, we first cross-referenced the sample of periodic sources in \cite{mor11} with the SED classification database subsequently presented by \cite{meg12}. We thus identified 9 possible sources as potential Class~I periodic variables. We subsequently discarded two of these sources, as their SEDs are clearly inconsistent with Class~I sources. Instead, they are Class~II or III sources that were misclassified in \cite{meg12}. The remaining 7 objects, on which we henceforth focus, are listed in Table\,\ref{tab:sample}. 
To ensure that these 7 objects are truly representative of Class I sources, we consulted the color-color diagrams of \cite{meg12}, and placed our objects in them using photometry data. 
%One such diagram is presented in 
The one diagram in which we can display all our objects is shown as
Figure\,\ref{fig:spitzer_ccd}. The near- and mid-infrared colors of our target stars span the range of
Class\,I sources as identified by \cite{meg12}. In the absence of systematic differences in color, we conclude that these sources are indeed representative of the larger population of Class\,I sources in Orion.

\begin{figure*}
\plottwo{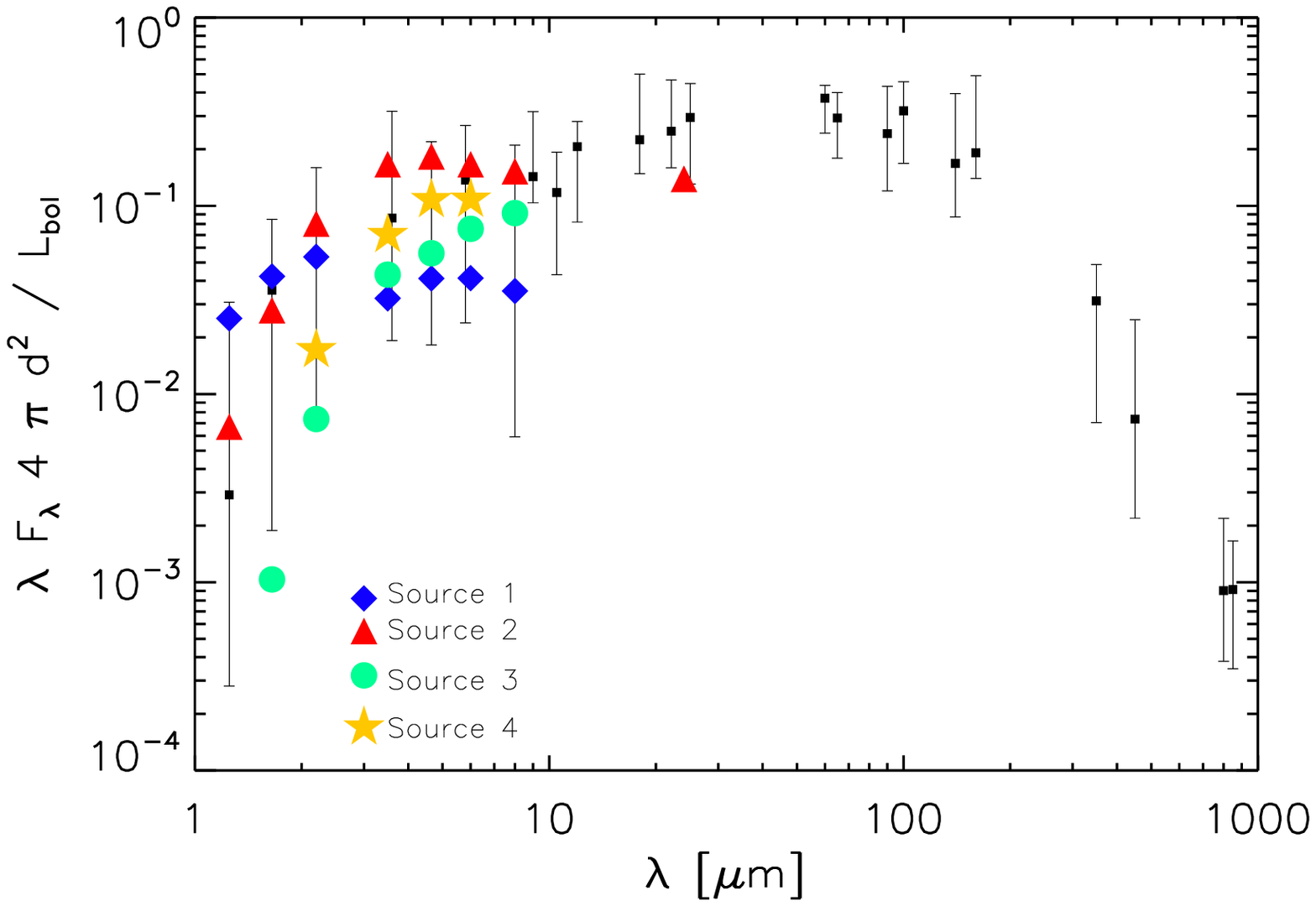}{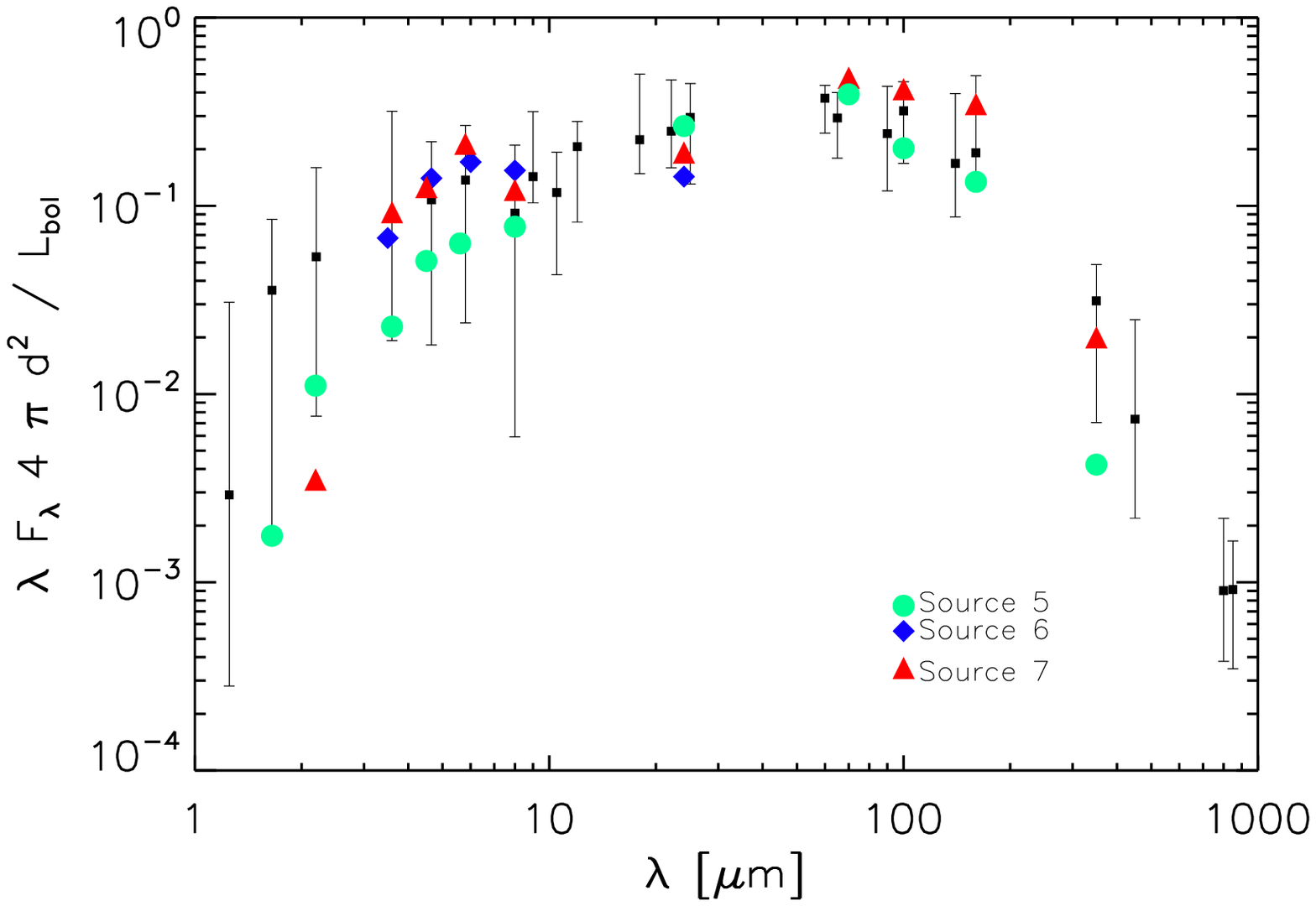}
\caption{SEDs of all 7 objects in our sample (colored symbols), superimposed over a template SED derived from Taurus Class~I sources (see Appendix\,\ref{appendix}).\label{fig:sed_data}. Sources 5 and 7 have additional points as they were included in the HOPS survey \citep{fur16}.}
\end{figure*}

Our next step was to construct more complete SEDs, using both 2MASS $JHK$ photometry \citep{skr06} and the {\it Spitzer} photometry published in \cite{meg12}. Additional photometry data for sources 5 and 7 were obtained from the HOPS survery in \citep{fur16}. The SEDs of all 7 sources are shown in Figure\,\ref{fig:sed_data}. In all cases, the SEDs are appropriately identified as Class~I, since the flux density either steadily increases towards longer wavelengths or is flat across the near- to mid-infrared regime; this behavior is also consistent with the Class~I template we have assembled (recall Figs.~1 and 2). 

We also inspected the 2MASS, {\it Spitzer} and WISE \citep{wri10} images of each target to search for potential source confusion. Specifically, we wanted to test the possibility that some of the Class I periodic sources are actually composed of two unrelated objects that lie in close proximity in the sky, a periodically variable star (possibly a Class II source in Orion) and an infrared-bright source. Given its relatively low angular resolution (several arcsec, depending on wavelength), {\it Spitzer} would not be able to resolve the two objects, and would thus erroneously assign the periodicity and infrared excess to the same object. All targets were detected in the 2MASS images and none shows any sign of confusion. It is still possible that some binary systems, such as a Class II source and a so-called "infrared companion" at close separation \citep{kor97}, could still escape inspection in this manner. However, the low occurrence rate of such systems indicates that they are unlikely to be present in all cases. In addition, Sources\,2, 6 and 7 have been assessed to be single in near-infrared HST observations by \cite{kou16}. We therefore conclude that the two key properties defining our sample -- periodicity and Class I nature -- are truly associated with a single object in each source.

\begin{table}
\centering
\caption{Periodic Class I sources analyzed in this study from \citet{mor11}. The fourth column gives the amplitude $\alpha$ of the photometric variations associated with the spot and the fifth column indicates the reduced chi squared value for the best-fit model (see Section\,\ref{sec:spot_fit}).
\label{tab:sample}}
\begin{tabular}{clccc}
\hline
\hline
Source & Name                & P [d] & $\alpha$ [mag]  & $\chi^2_{\mathrm{red}}$  \\
\hline
1 & ISOY J053500.22-052409.2     & 7.29               & 0.39         &  3.12         \\ 
2 & ISOY J053516.33-052932.7     & 16.79             & -0.19    &       2.72        \\ 
3 & ISOY J053520.72-051926.5       & 22.21          & -0.06       &     4.09     \\
4 & ISOY J053522.10-051857.6     & 5.78            & -0.08      &       2.27       \\
5 & ISOY J053524.73-051030.1     & 4.37            & -0.08        &     1.91     \\ 
6 & ISOY J053525.18-050509.3       & 5.64            & 0.13     &       1.59    \\
7 & ISOY J053528.19-050341.2       & 23.52          & -0.23       &     2.66       \\
\hline 
\end{tabular}
\end{table}

\subsection{Fitting the light curves}\label{sec:spot_fit}

\begin{figure}
\plotone{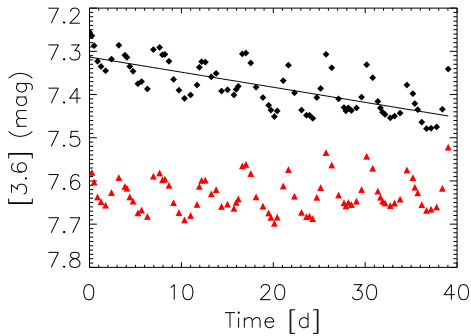}
    \caption{Original (black diamonds) and detrended (red triangles) light curve for Source~5. The black solid line is a linear fit to the original light curve, which is subtracted to obtain the detrended light curve. An arbitrary offset has been added to the detrended light curve for clarity. 
    }
    \label{fig:detrend}
\end{figure}

We retrieved the photometric time series from \cite{mor11} for the 7 Class I periodic sources. Visual inspection revealed that, in at least several cases, the quasi-periodic variations that were initially identified through periodogram analysis occur within longer-timescale fluctuations. \cite{mor11} suggest that these drifts reflect slow changes in the accretion rate or else self-shadowing in a warped disk. We remain agnostic about the nature of these long timescale fluctuations, but attempt to remove them in order to focus our analysis on the faster, manifestly periodic part of the signal. To this end, we subtract from the time series a linear trend obtained through least-squares fitting (see Fig.\,\ref{fig:detrend}). The amplitude of this linear variation over the 40\,d of the {\it Spitzer} campaign ranges from 0.03 to 0.17\,mag. 

\begin{figure*}
\plottwo{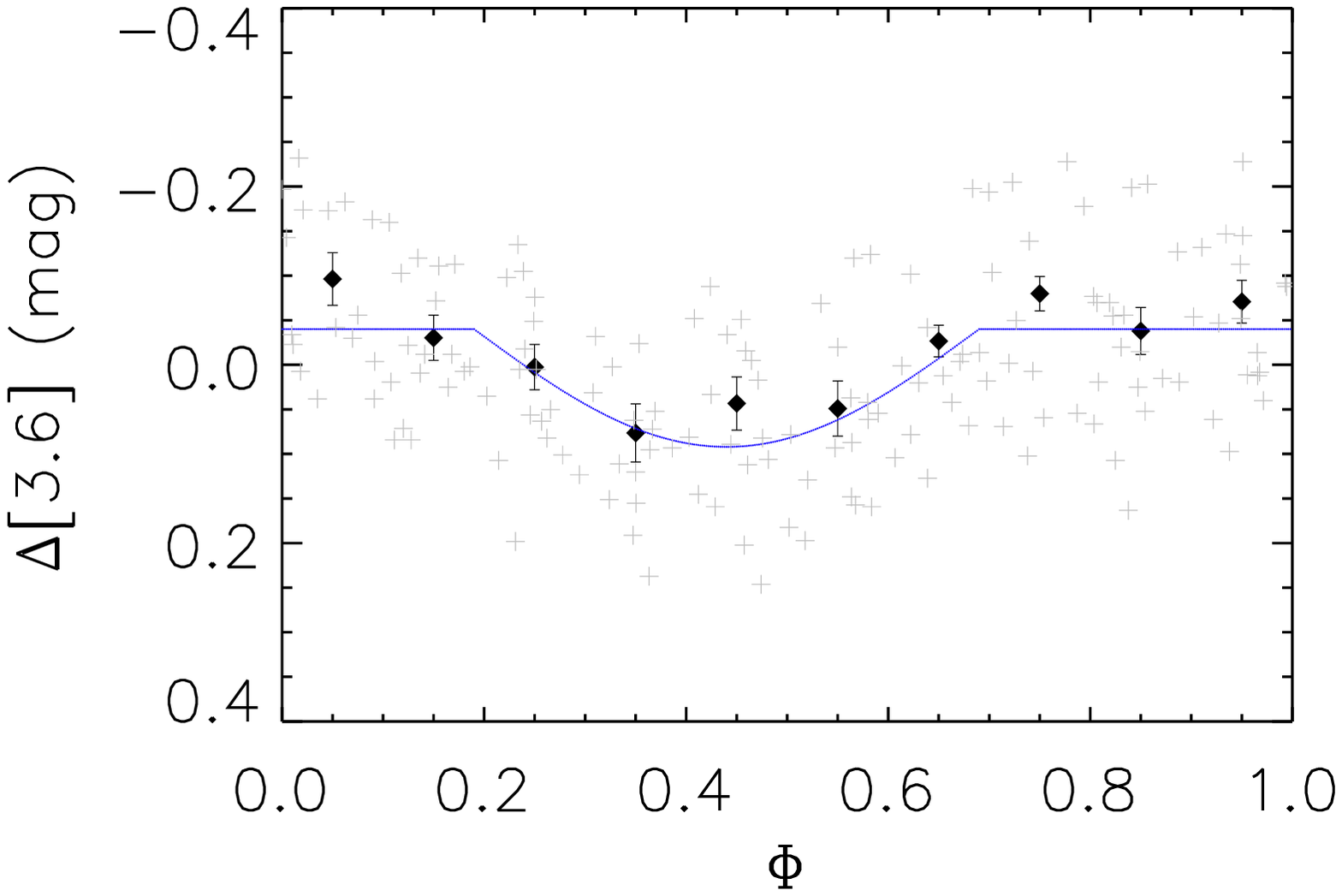}{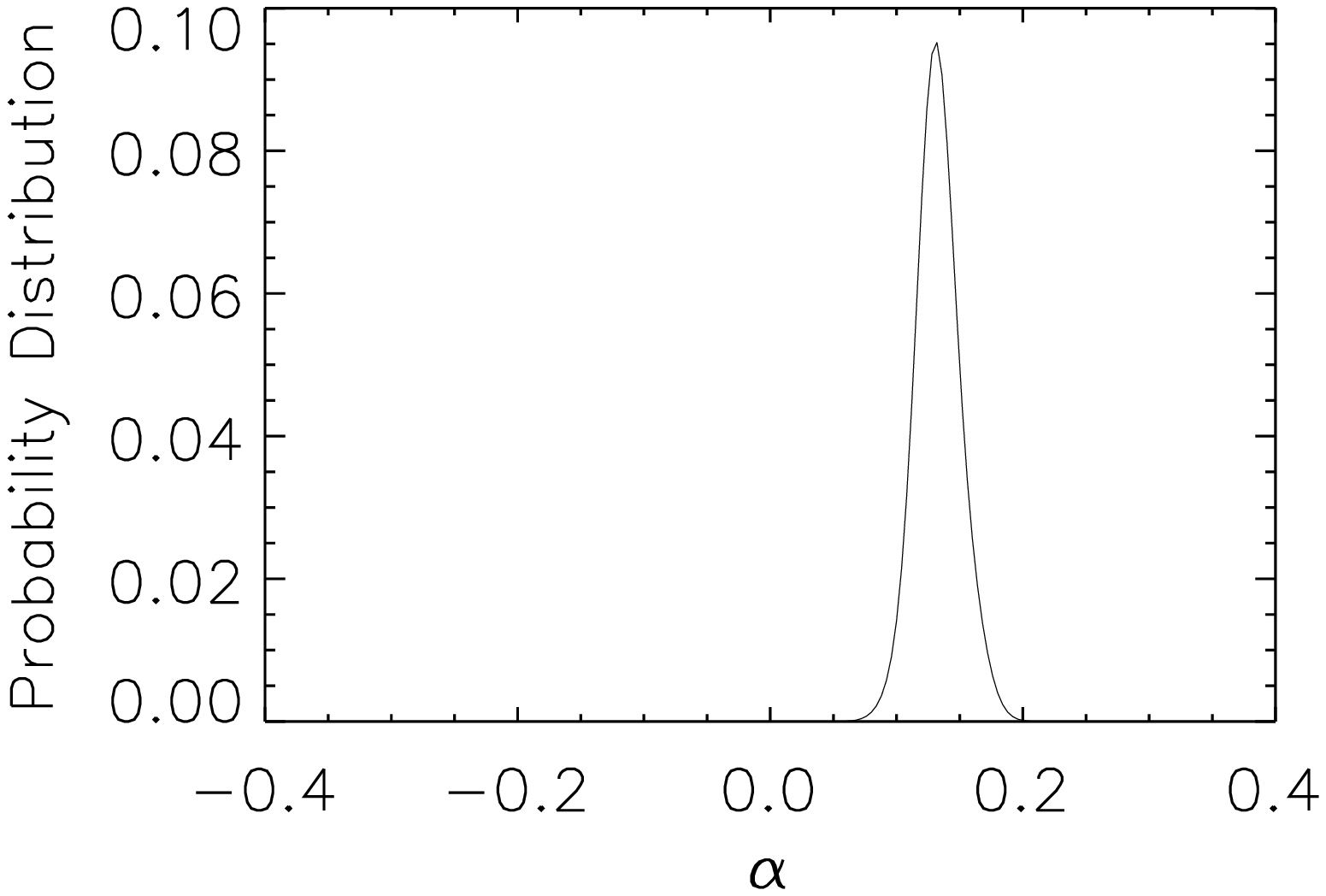}
    \caption{{\it Left:} Phased light curve for Source~6, an example of periodic dimming. The gray crosses and black diamonds represent the detrended and binned light curves, respectively, whereas the solid blue curve is the best fitting "half-cosine" model. The binned light curve is shown for illustration only; the model is fit to the individual, detrended datapoints. The phase has been shifted so that the spot feature is centered around $\phi \approx 0.5$. {\it Right:} Posterior probability distribution for the amplitude of the half-cosine model; the positive values of $\alpha$ indicates that the star gets temporarily fainter as the spot comes into view. Equivalent figures for the other sources are available in the electronic version of the journal.} 
    \label{fig:lc_cold}
\end{figure*}

Once the photometric time series were detrended, we transformed them into phase-folded light curves using the periods proposed by \cite{mor11}. The scatter in the phased light curves is usually much larger than the quoted photometric precision (0.02--0.04\,mag, typically), indicating that there is significant residual fluctuation from period to period even after detrending. Nonetheless, each light curve clearly departs from a flat line (see Fig.\,\ref{fig:lc_cold}), confirming the robustness of the original periodogram analysis.

We next assess whether the observed photometric variations can be ascribed to spots on the stellar surface. We focus on the 3.6\,$\mu$m timeseries, but note that the same analysis of the 4.5\,$\mu$m data lead to very similar results, albeit with an amplitude of the periodic signal that is slightly larger. As we will see, a simple model of a rotating star with a single spot on its surface does well to explain the observed, periodic variations. However, we acknowledge
that the true source of variations could be multiple spots, or persistent disk features that somehow produce similar signals.

We assume that the spot covers a relatively small fraction of the stellar surface that does not overlap with the polar axis. We further assume that the spot itself can be idealized as an area of uniform brightness. The expected photometric signal introduced by the spot is then directly proportional to its projected area on the stellar surface. If the spot is small enough and disappears from the observer's view during each rotation period, this projected area is proportional to the cosine of the relative rotational phase $\phi - \phi_0$, where $\phi_0$ is the phase at which the spot center crosses the stellar meridian. This idealized model ignores limb darkening, temperature gradients within spots, and the effects of large and/or multiple stellar spots that are commonly associated with young stellar objects \citep[e.g., as seen in Doppler imaging;][]{ske08, ske09}. Nonetheless, we believe this model is sufficient for our limited goals --  to assess whether the photometric variability is plausibly due to a rotating spot, and to provide a quantitative measure of the amplitude of that variability. 

We thus proceed to perform a least-squares fit to each object's light curve, employing three parameters:
\begin{equation}
[3.6](\phi) =
\begin{cases}
[3.6]_0\,+\, \alpha\,{\rm cos}\,[2\pi(\phi-\phi_0)] & \text{if $|\phi-\phi_0| \leq 0.25$} \\
[3.6]_0  & \text{otherwise.}
\end{cases}
\end{equation}        
Here, $[3.6]_0$ is the apparent 3.6~${\mu}m$ magnitude of the star when the spot is not in sight and $\alpha$ represents the amplitude of the spot-induced signal. In principle, the sign of $\alpha$ indicates the nature of the spot: a cold spot which makes the star temporarily fainter, corresponds to a positive $\alpha$. Conversely, a hot spot induced by an accretion shock onto the stellar surface produces a negative $\alpha$.

We first explored parameter space through a fine sampling of 
$[3.6]_0$, $\alpha$, and $\phi_0$. When we considered only the quoted photometric errors in the data, even the best-fit model curves for each source had a relatively large $\chi^2$. The source of the problem was the residual, period-to-period fluctuations mentioned previously. To address this issue, we subtract the best-fit curve from the empirical data for each source, then calculate the "variability scatter" as the dispersion of the residuals. We quadratically add this scatter to the nominal photometric error to produce an "effective" uncertainty. This procedure assumes that the two errors are both random and uncorrelated with each other. 

We then repeat our exploration of parameter space, but now using the enhanced, effective errors. Ideally, the resulting $\chi^2_{\rm red}$ of each best-fit model should be close to unity. However, as seen in Tables~2 and B1, the actual  
$\chi^2_{\rm red}$-values are still generally larger. We interpret this finding as an indication that the fluctuations are in fact correlated to some degree. In any case, we judge the model fits to be adequate for our purposes. Since $[3.6]_0$ and $\phi_0$ are not the focus of this study, we have marginalized over both these quantities to produce posterior probability distributions of $\alpha$ for each source. One representative light curve and $\alpha$-distribution is in Figures~\ref{fig:lc_cold}.

In all cases, the posterior distribution has one distinct peak. The corresponding $\alpha$-value of that peak is the amplitude of the best-fitting model; these values are listed in Table\,\ref{tab:sample}. 
The amplitudes range from 0.05 to 0.39\,mag; weaker variations would probably remain undetectable given the {\it Spitzer} photometric precision and the intrinsic scatter around perfectly periodic signals displayed by our targets' light curves. We note that in all cases models with $\alpha = 0$ are excluded by the probability distribution, confirming that there is a significant periodic signal in the light curve. Ostensibly, the prevalence of negative values of $\alpha$ for the best-fitting models suggests that the variability of most (but possibly not all) periodic Class\,I sources is due to  accretion-generated hot spots.

\subsection{Comparison with Class~II and III sources}\label{subsec:per_II_III}

It is instructive to compare the amplitudes of the periodically variable Class~I sources in Orion with those of their Class~II and III counterparts. For this purpose, we consider the sample of 143 variables whose periods were established by
\citet[][see their Table~4]{mor11}.
We assigned an evolutionary class to each object by locating it in the catalog of \citet{meg12}. If an object was not classified by \citeauthor{meg12} as a ``protostar" (Class~I) or ``disk-bearing" (Class~II), it was assumed to be Class~III. Of the 143 variables, 7 are Class~I, as previously noted, 63 are Class~II, and 76 are Class~III.
Taking each object's light curve from \citet{mor11}, we then performed our least-squares fit to establish the amplitude $\alpha$, in the same manner as for Class~I sources. In almost all cases, the posterior distribution of $\alpha$-values shows one distinct peak. The distribution is more bimodal for a handful of sources with intrinsically weak variability. We list in Table~B.1 the amplitude, period, and evolutionary class of all 143 objects. 

While we were able to find least-squares fits to all Class~II and III light curves, the limitations of our simple, half-cosine model became apparent with this larger sample. For instance, two of the Class~II sources, ISOY\,053605.95-050041.2 and ISOY\,J053518.03-052205.4, are eclipsing binaries \citep{mor11} and thus not well described by our model. 
 
\begin{figure*}
    \plottwo{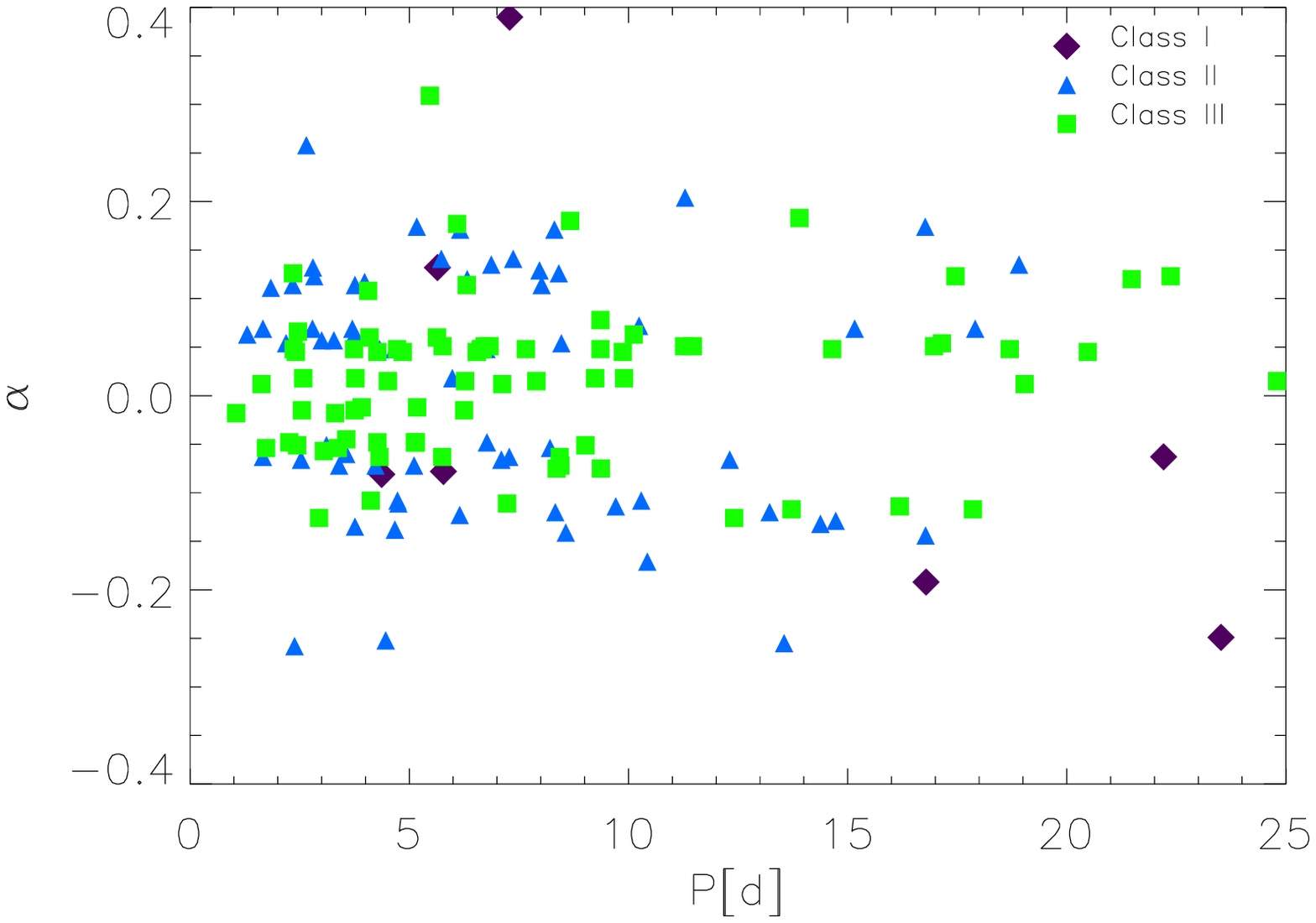}{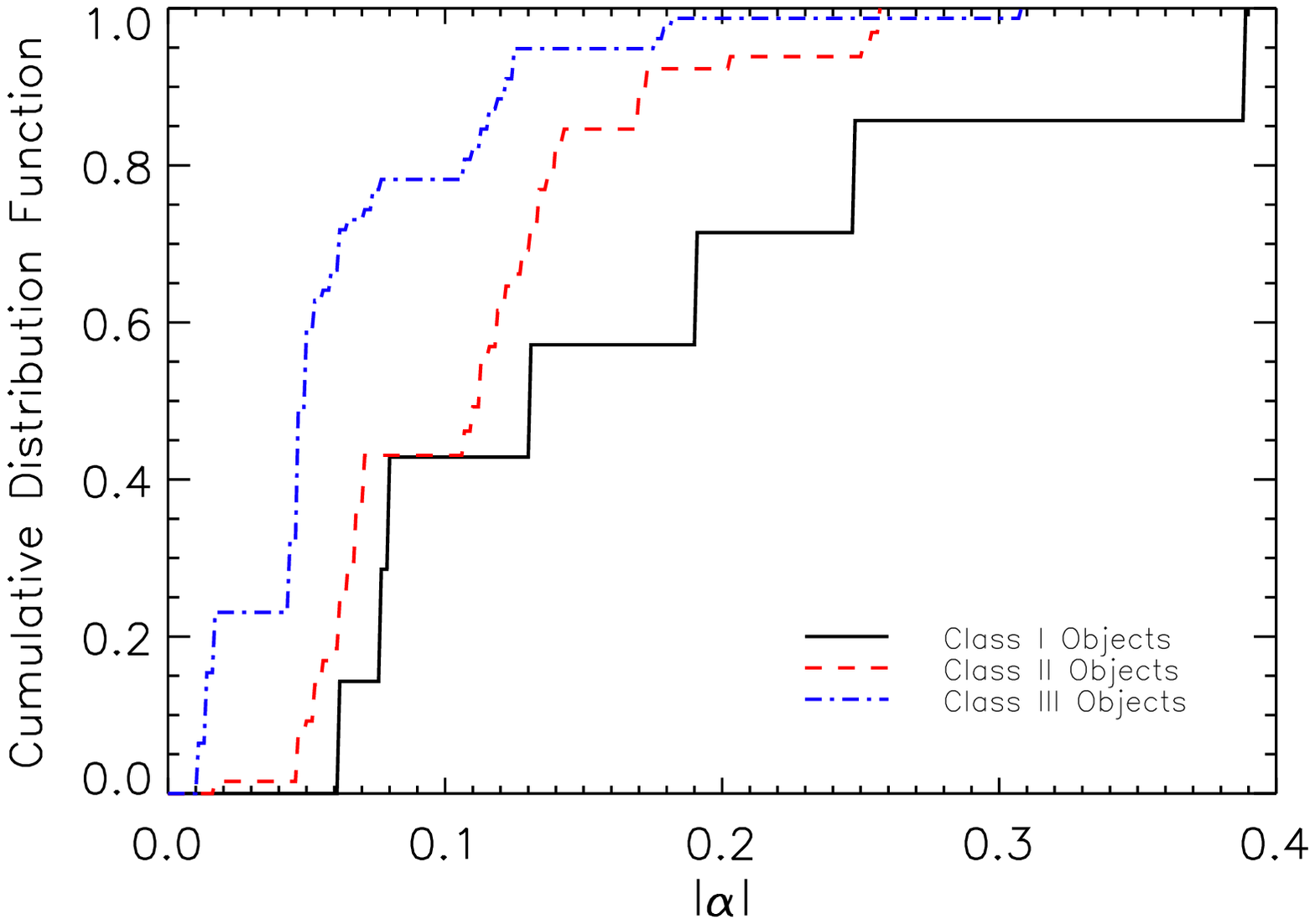}
    \caption{\textit{Left:} Variability amplitude as a function of period for all Orion periodic sources. Class\,I, II and III are indicated by purple diamonds, blue triangles and green squares, respectively.  \textit{Right:} Cumulative histograms of the variability amplitude for each of the three types of sources. The solid black line denotes Class\,I, the dashed red line Class\, II, and the dot-dashed blue line Class\, III.} 
    
    \label{fig:alpha_scatter}
\end{figure*}

In addition, we would expect that Class~III sources, corresponding to non-accreting T~Tauri stars, would have only cold spots (\hbox{$\alpha\,>\,0$}). However, our fitting procedure finds that many have \hbox{$\alpha\,<\,0$}. This apparent incongruity likely stems from the presence of multiple spots and departures from pure sinusoidal behavior even in the single-spot case. In summary, the {\it algebraic sign} of $\alpha$ cannot be interpreted literally. We maintain, however, that the {\it magnitude} of $\alpha$ still signifies quantitatively the degree of variability. 
Pursuing this line of reasoning, we find that the variability of Class~I sources is much more similar to Class~II objects than to those of Class~III. Specifically, Class~III sources generally have small values of $|\alpha|$, while Class~I and II sources share a similar range in this quantity. The median values of $|\alpha|$ are 0.13, 0.11, 0.05 for Class\,I, II and III, respectively.
The left panel of Figure\,\ref{fig:alpha_scatter},
which plots $\alpha$ against the period $P$ for all three classes, shows the tendency. To confirm this impression more quantitatively, we performed a Kolmogorov-Smirnov test on the cumulative distributions of $|\alpha|$ for the three samples (see right~panel of Fig. \ref{fig:alpha_scatter}).
Table\,\ref{tab:ks_test} lists the resulting values of $P_{KS}$ for the various pairs of samples. From these values, we can exclude the possibility that the Class~I and III amplitudes were drawn from a single, underlying distribution. Similarly, the Class~II and III sources do not share the same \hbox{$|\alpha|$-distribution}. On the other hand, periodic Class~I and II sources have consistent amplitudes, indicating that the underlying physical objects could be identical. Previously, \citet{wol15} found a similar result through observations of the GGD12-15 star-forming region (see their Fig.~7). We ran a similar Kolmogorov-Smirnov test on the distribution of periods for each class of object. The results are shown in Table\, \ref{tab:ks_test}, and do not indicate any significant difference in the distribution of periods between any two classes. 

\begin{table}
    \centering
    \begin{tabular}{lll}
    \hline
    \hline
    Object Categories     &  \hspace{7mm} $ P_{KS}$ &\\
     & $| \alpha |$ & P[d]\\
    \hline
     Class I and Class II   & $0.33$ & $0.21$\\
     Class I and Class III & $0.003$ & $0.39$\\
     Class II and Class III & $3.1 \; 10^{-5}$& $0.83$\\
     \hline
    \end{tabular}
    \caption{$P_{KS}$ values for each combination of object as obtained from a Kolmogorov-Smirnov test on their $| \alpha |$ and period distributions. }
    \label{tab:ks_test}
\end{table}

\subsection{Envelope and disk models for Class~I light curves}

We have seen that both TSC and power-law (\hbox{$\gamma\,=\,1$}) envelopes, in conjunction with circumstellar disks, can successfully reproduce the typical SED of a Class~I source. We next examine whether endowing the central star with a rotating spot can yield a light curve at 3.6~$\mu$m that broadly matches the Orion observations. Following the current modeling of accreting, Class~II sources, we consider the spot to be an optically thick, $10^4$~K layer of gas at the stellar surface
\citep[e.g.,][]{har16}. We place the spot at the star's equator, and  select its surface area to be about 3\% of the stellar surface so that the relative modulation at 3.6~$\mu$m from the star (with a disk but without any envelope) is 15~percent. This amplitude is motivated by the observations of periodic Class~II variables in Orion (recall Fig.~\ref{fig:alpha_scatter}). To generate our synthetic light curves for this asymmetric system, we use the 3D version of MCFOST. As in the previous, SED calculations, we set the observer's line of sight at an inclination angle of $57^\circ$.

\begin{figure*}
\plottwo{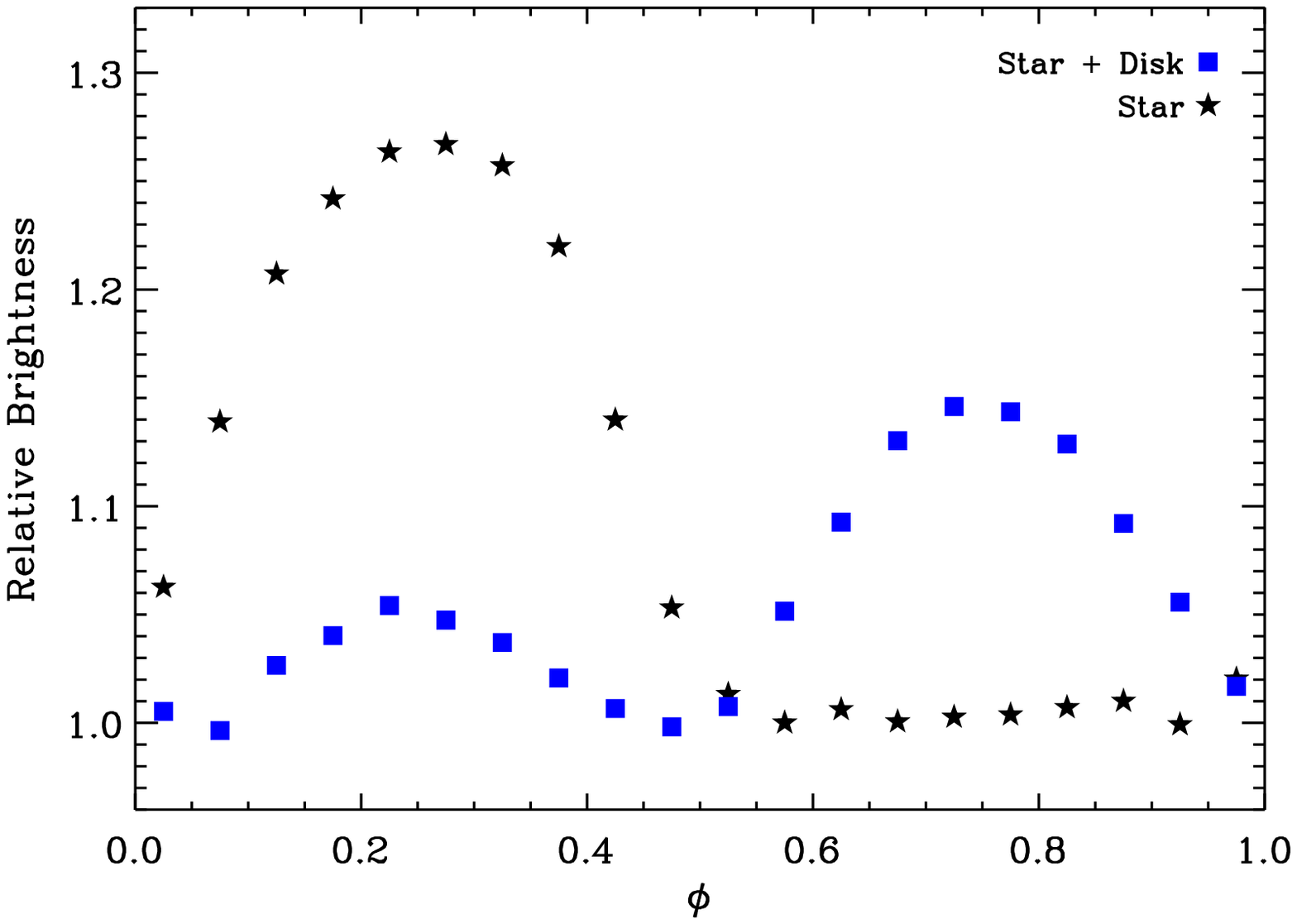}{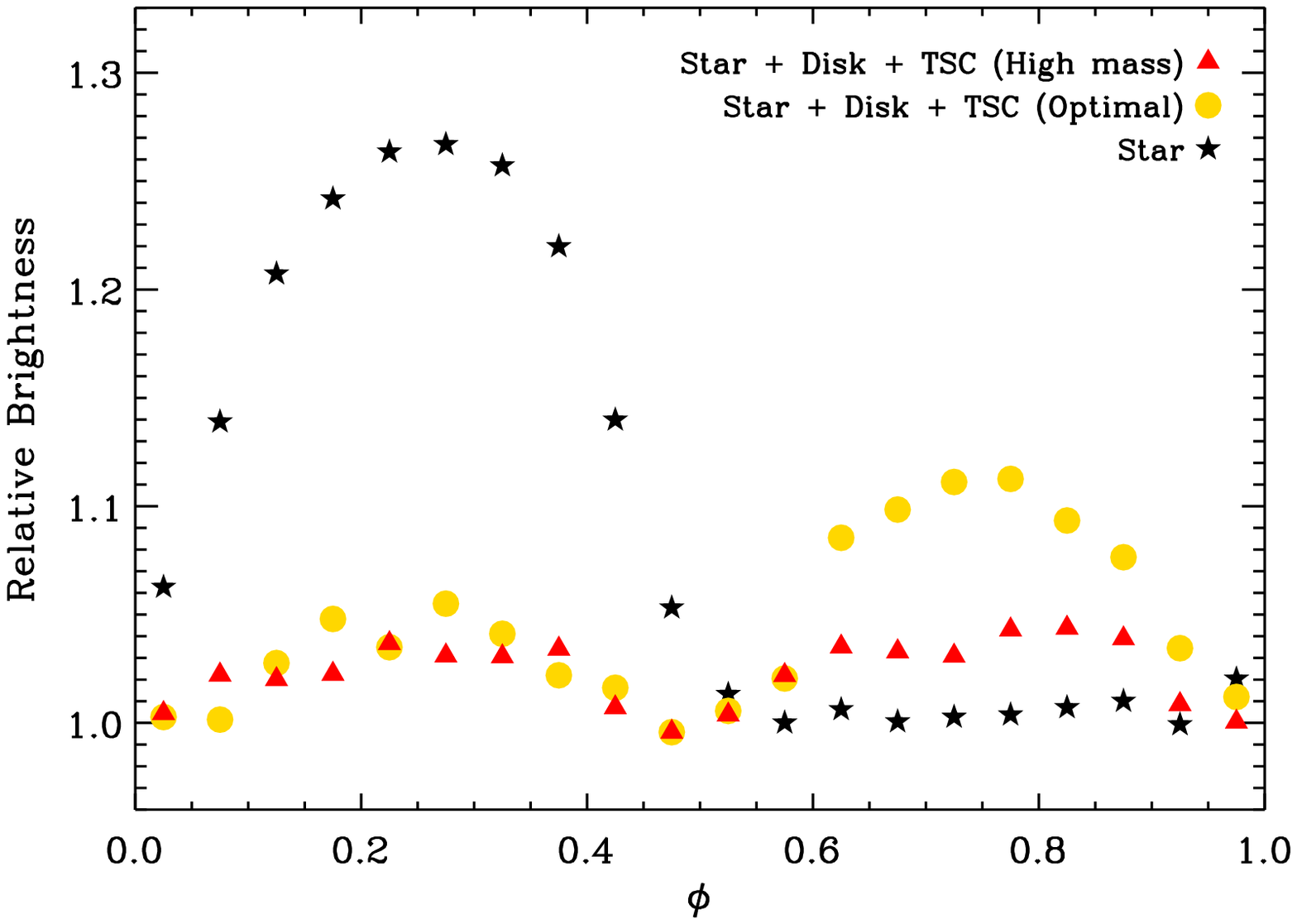}
\plottwo{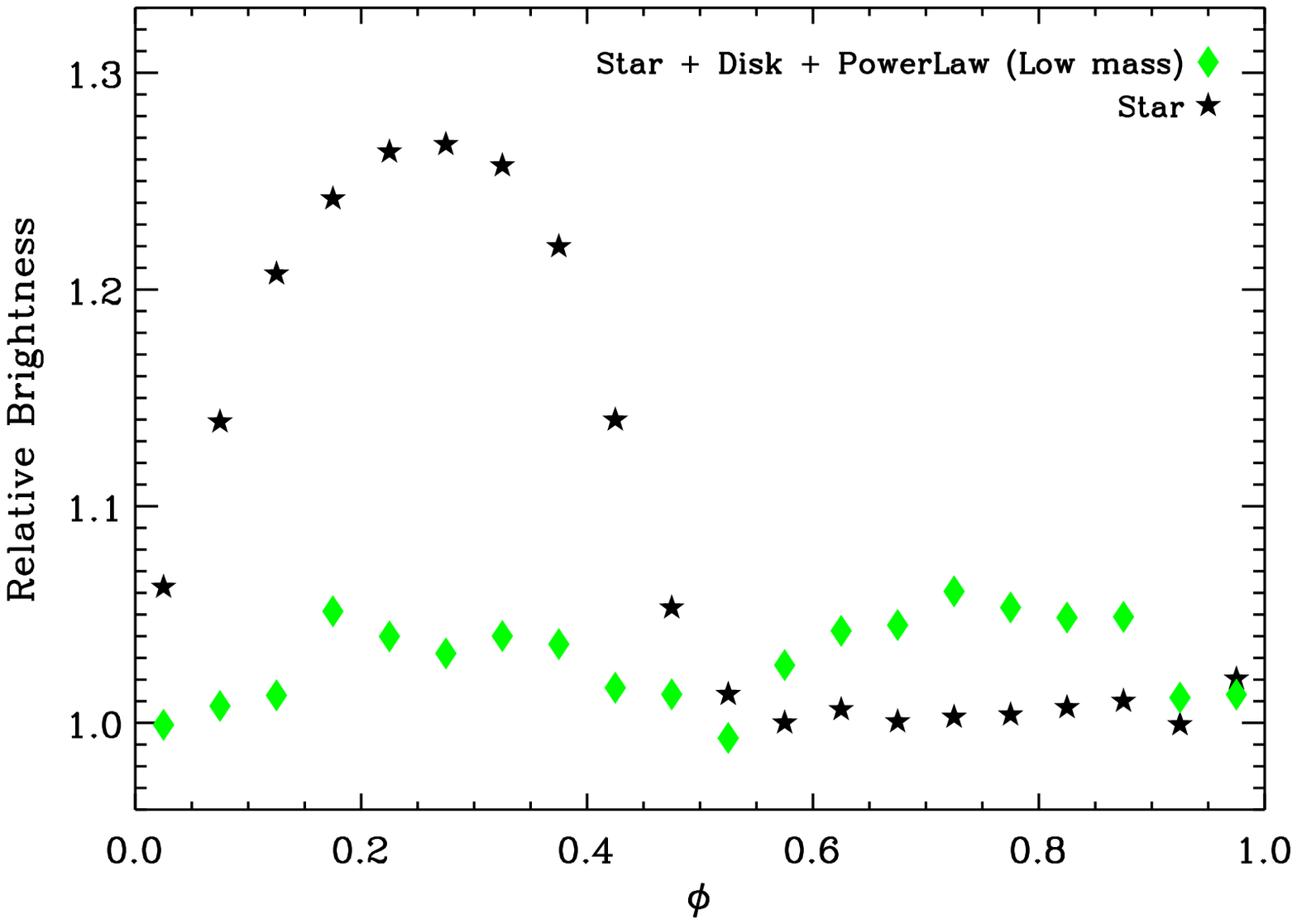}{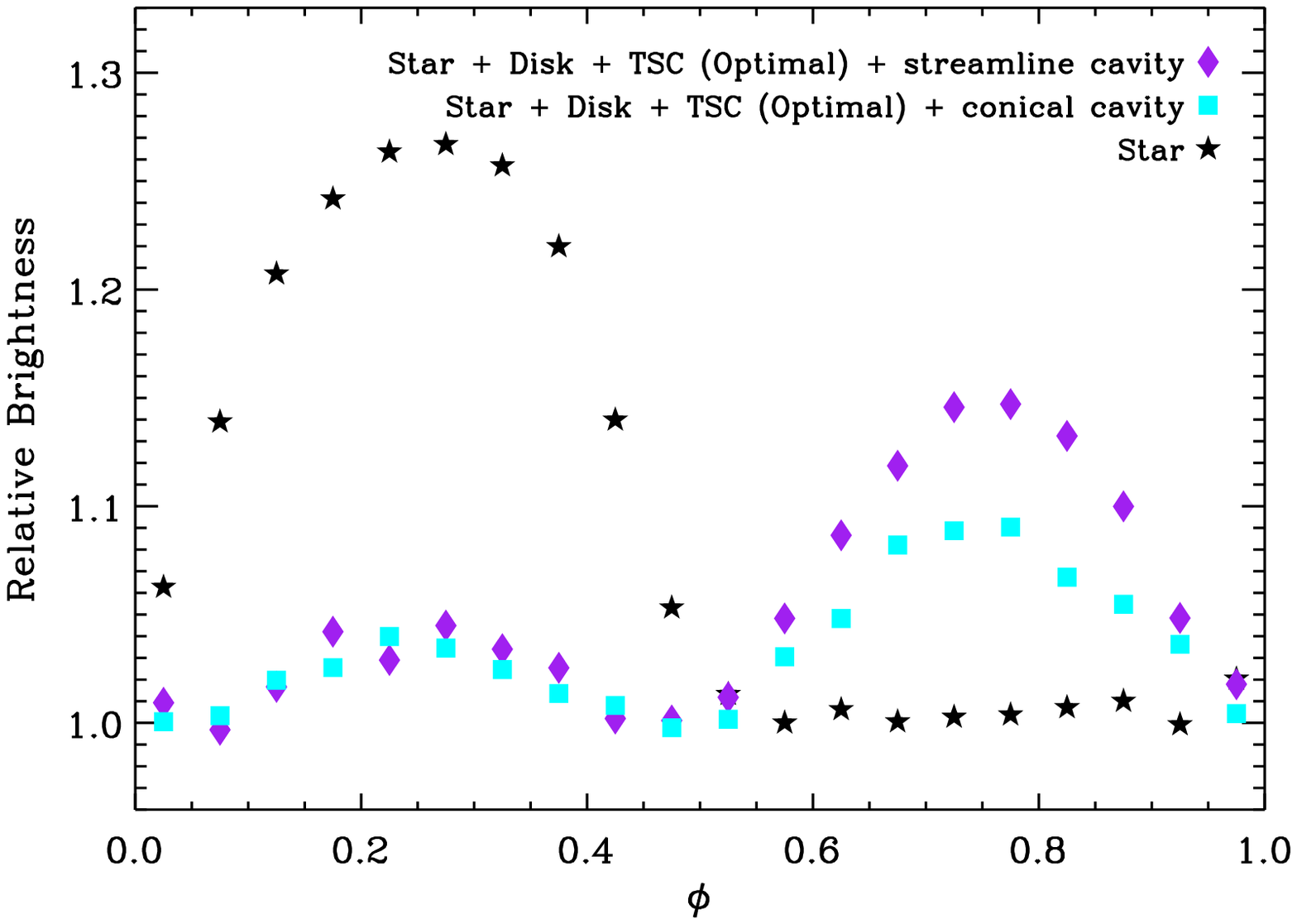}
\caption{Model light curves for some models discussed in this study. The light curve for an isolated spotted star, indicated by black star symbols, is repeated in each panel. The light curve for the same star with a circumstellar disk is shown in the top left panel, whereas models of Class\,I sources using the TSC and powerlaw envelope models are displayed in the top right and bottom left panels, respectively. Finally, the bottom right panel illustrates the effect of introducing a polar cavity in the optimal TSC model.
\label{fig:model_lightcurve}}
\end{figure*}

We examine first the light curve generated by a spotted star hosting a disk but without any envelope, i.e., a Class~II source. In this case, thermal emission from the inner region of the disk overwhelms the stellar photosphere by a ratio of 4:1 at 3.6~$\mu$m. In addition, the area of the inner disk that directly faces the stellar hot spot is locally overheated, hence brighter. As a consequence, the light curve of such a system displays, in each period, two maxima in opposition of phase (see top panel in Figure\,\ref{fig:model_lightcurve}). The first maximum, in phase with the stellar spot, has a much reduced amplitude compared to the diskless case. A second maximum, due to local overheating of the disk by the hot spot, is observed in opposition of phase with the stellar spot, when this disk region is in the observer's direct view. 

We note parenthetically that the optical light curve of the star-disk system would exhibit a single maximum in phase with the spot, since disk emission is negligible in this regime. \citet{cod14} have observed this contrast between near-infrared and optical light curves in Class~II sources, while \cite{kes16} have found a similar effect in simulations.

We now proceed to compute the light curve of a system containing a spotted star, a circumstellar disk and an envelope, i.e. a full-fledged Class\,I system. Results from this analysis are shown in Figure\,\ref{fig:model_lightcurve} (middle and bottom panels). 

Starting with the family of TSC models, we find that both the low-mass and best-fit models have only a very modest effect on the star-disk light curve consistent with the fact that these envelopes are optically thin along the line of sight at 3.6\,$\mu$m. The high-mass model, on the other hand, is sufficiently opaque to redistribute azimuthally the excess emission induced by the hot spot at the disk's inner edge. As a result, the light curve only shows minor ($<5\%$) variations that would only be marginally detectable through {\it Spitzer} monitoring. The $\gamma=1$ power law family of models is overall slightly more optically thick. The low-mass model is already sufficiently opaque to reduce the variability amplitude to $\approx8\%$, still consistent with observations. On the other hand, the best-fit and high-mass model completely damp out any photometric variability.

It is well known that Class\,I sources drive polar outflows and jets that are believed to be responsible for the presence of polar cavities in their surrounding envelopes. This offers a different route for photons to escape the system. Besides their already studied effects on the SED \citep[e.g.,][]{whi03, fur16}, one may wonder whether these cavities could alter significantly the lightcurve induced by a spotted central star. To test this possibility, we computed the lightcurves for additional models using our optimal TSC envelope structure but with a polar cavity carved out. The exact shapes of these cavities are not well established observationally, so we adopted two different morphologies, to explore the range of possibilities. Following \cite{whi03}, the first cavity shape we adopt is defined by the infall streamlines within the TSC model. The second cavity model has a simple conical shape. In both cases, we select an outer opening angle of 20\degr, in line with scattered light images of Class\,I sources \citep{gra10}. Other curved cavity shapes have been explored in the past \citep{eis05, fur08}, but they are bracketed by our two options. In all cases, our chosen line-of-sight is at a higher inclination than the cavity opening. Thus, all the received light is reprocessed by the envelope. With too small a viewing angle, some light would travel directly from the star to the observer, and the resulting SED would be much more elevated at short wavelengths than in typical Class~I sources.

The resulting lightcurves, presented in the bottom right panel of  Figure\,\ref{fig:model_lightcurve} show that the effects of the cavity are modest, with the periodic signal maintaining a similar amplitude with and without a cavity. While photons can escape more easily from the system, most of them are not directed to an observer whose line of sight intercepts most of the envelope. Only the original signal, mildly attenuated by opacity through the envelope as we have shown before, reaches the observer. Finally, we note that the conical cavity reduces the amplitude of the signal more than the streamline cavity. In the latter case, the bounding streamline intersects the midplane 13~au from the central star. Since the region inside is evacuated, the optical depth along the observer's line of sight is diminished.

From this exploration, we conclude that some, but not all, of the SED-acceptable models of envelopes for Class\,I sources are compatible with the %$\geq10\%$ 
variations observed in the Orion Class\,I periodic sources. Specifically, the simultaneous constraints imposed by the objects' SED and their amplitude of variability imply that their envelopes must have $\tau_{3.6}$ of approximately 0.6--1.0 along the observer's line of sight. More opaque envelopes would reduce the photometric effects of the spotted star below their observed amplitude, while less opaque envelopes would result in too low a circumstellar extinction and thus too excessive near-infrared emission reaching the observer. For the TSC and $\gamma=1$ powerlaw envelope models that we have considered in this study, the combined constraints from SEDs and light curves favor a total envelope mass of 0.025--0.05\,$M_\odot$, i.e., an order of magnitude less than that of the central star.

% ==============================================================================================
\section{Discussion}
\label{sec:discus}

We now return to the fundamental question at the heart of our study, the physical nature of Class~I sources. By studying the periodically variable ones, we first confirm that the central star in every case must be embedded in a three-dimensional circumstellar envelope. Unlike Class~0 sources, however, this structure must be of modest mass with respect to the star, 5 to 10\% the mass of the stellar mass for our best models. Only then can the photometric variability, which certainly originates in the innermost region (given its timescale) be detected by an outside observer. An envelope that is too optically thick erases any photometric signal, both by reprocessing the emitted light to much longer wavelengths, and by diffusively spreading any temperature anisotropy. 

The central source that illuminates this moderately opaque envelope has photometric behavior that is consistent with Class~II sources, which usually have disks, but not substantial envelopes. In other words, it appears that at least the periodically variable Class~I sources in Orion are otherwise normal pre-main-sequence stars that happen to be embedded in moderately opaque envelopes. This picture corroborates previous findings that the spectroscopically determined effective temperatures, rotation rates, veiling, and line emission are similar in Class~I and Class~II sources \citep[e.g.,][]{whi04}.

While mutually consistent, the dual constraints on the optical depth of Class\,I sources imposed by the SEDs and light curves ($0.5 \lesssim \tau_{3.6}^{\mathrm{env}} \lesssim 4$ and $\tau_{3.6}^{\mathrm{env}} \lesssim 1$, respectively) only partially overlap. Thus, the sample of Class\,I sources that display a periodic photometric signal is likely biased towards those with less opaque envelopes. Class\,I sources with non-periodic light curves could be either (slightly) more embedded objects, and/or aperiodic; the latter case is common among Class\,II sources \citep{mor11}.  Either way, we speculate, based on our analysis of the Orion periodic variables, that all Class\,I sources harbor normal pre-main sequence stars.

As for the internal structure of the envelope itself, our study provides only broad constraints. The widely used TSC envelope does yield a synthetic SED that roughly matches those of Class~I sources. The fact has long been known \citep[e.g.,][]{whi03}, but two caveats apply. First, we find that the envelope mass can be no greater than about $0.05~M_\odot$ to preserve the periodic signal in the 3.6~$\mu$m light curve. Within the framework of the underlying TSC model, this minimum mass corresponds to an envelope infall rate of $1.3\times 10^{-6}\,M_\odot\,{\rm yr}^{-1}$. However, we emphasize again that current observations inform us only about the envelope structure, and not its dynamics, and that this infall rate is in reality a proxy for the total envelope mass. 

The geometric symmetry of Class~I envelopes is still uncertain. In this regard, we have shown that inclusion of an outflow cavity has little impact on the observed SED or near-infrared lightcurve. Note that, even without an outflow cavity, TSC models are not spherically symmetric, since the density is lowered along the cloud's rotation axis. However, even this asymmetry is not essential for the envelopes to satisfy the two requirements of matching a typical Class~I SED \citep[see, e.g.,][who modeled several Taurus Class\,I sources]{fur08},
and producing a satisfactory near-infrared light curve. We have illustrated this point by constructing much simpler envelope models that are spherically symmetric and have power-law density profiles, 
$\rho (r) \propto r^{-\gamma}$. Such models yield acceptable SEDs and light curves, as long as the envelope mass is relatively small and the power-law exponent $\gamma$ is roughly 1, the specific value we employed. Note that the azimuthally-averaged density profile in the TSC model approaches, at small radii, a power law with $\gamma = 0.5$. Interestingly, the first imaging studies for Class~0 envelopes have found that pure power-law envelopes models yield a better fit than TSC models \citep[see, e.g.,][]{tob15,per16}. 

More detailed study of Class~I envelope structure  will require spatially resolved mapping. Such observations need to be done at submillimeter or millimeter wavelengths, where the envelope is optically thin. In this regime, unfortunately, thermal emission from the cold outer regions of the circumstellar disk exceeds that from the envelope. Nonetheless, progress is being made. The PROSAC project used the SMA to disentangle envelope and disk masses in both Class~0 and I sources \citep{j09}.

If we are correct that the central stars in Class~I sources are pre-main-sequence objects, then many will have hot spots on their surfaces, like their Class~II counterparts. It is generally accepted that these are created by the transfer of mass from the inner regions of a circumstellar disk. The question then naturally arises as to whether a portion of the envelope is falling onto that disk to replenish it. Addressing this dynamical issue empirically will require spectral line observations of Class~I envelopes. Such studies have also begun, both as part of the PROSAC project mentioned earlier, and in targeted observations of individual sources \citep[e.g.,][]{L16}. So far, the emphasis has been on the dynamics near outflow cavities; we hope studies will broaden to include the more general issue of cloud collapse.

Returning to our own work, it will be important for the future to increase the sample size of periodic Class~I sources.
The YSOVAR group has detected periodic variables in other star-forming regions \citep{gun14, pop15, reb15, wol15}. It will be interesting to apply our analysis to these regions to expand the sample of Class\,I periodic sources.

To bolster the idea that the central objects are normal, pre-main-sequence objects, near-infrared spectroscopy would be helpful. Such observations would not only yield estimates of the effective temperature of the central objects, but also of the accretion rate on the stellar surface via continuum veiling and hydrogen line emission (e.g., Pa$\beta$, Br$\gamma$). Finally, Zeeman observations could reveal the magnetic field strength associated with these sources, to see if magnetic forces play a role in envelope structure. \\ \\ 

\section*{Acknowledgements}

The authors are grateful to Keivan Stassun for early discussions about this project, to Bill Herbst for input on the periodic variability of Class~II and III sources, and to Christophe Pinte for his help in implementing and testing the addition of stellar spots in MCFOST. This research has made use of the NASA/ IPAC Infrared Science Archive, which is operated by the Jet Propulsion Laboratory, California Institute of Technology, under contract with the National Aeronautics and Space Administration. This publication makes use of data products from the Two Micron All Sky Survey, which is a joint project of the University of Massachusetts and the Infrared Processing and Analysis Center/California Institute of Technology, funded by the National Aeronautics and Space Administration and the National Science Foundation.
% ==============================================================================================

%%%%%%%%%%%%%%%%%%%%%%%%%%%%%%%%%%%%%%%%%%%%%%%%%%



%\bibliographystyle{mnras}


%%%%%%%%%%%%%%%%%%%%%%%%%%%%%%%%%%%%%%%%%%%%%%%%%%

%%%%%%%%%%%%%%%%% APPENDICES %%%%%%%%%%%%%%%%%%%%%

\appendix

\section{A Class I template SED}
\label{appendix}

It is necessary to assemble the observed SED of a Class I source in order to evaluate which of our model SEDs are representative of such objects. The SED of the targets studied in this work only extends out to 24\,$\mu$m (and often only 8\,$\mu$m). While this is sufficient to assess their embedded nature through their near- to mid-infrared spectral index, the model SEDs must be compared over a much broader range of wavelengths, spanning from the optical/near-infrared to the millimeter regime. Furthermore, there is a rich diversity of Class I SEDs, possibly due to viewing angle effects and an intrinsic diversity in envelope size and mass. To draw a representative ("template") SED, it is therefore necessary to consider a larger sample of Class I sources and to use their average and dispersion. Similar efforts have been conducted for Class II sources in the past \citep[e.g.,][]{mat13}.

To ensure the richest and most uniform dataset possible, we have focused on the population of Class I sources in the nearby Taurus-Auriga star-forming region, arguably the best-studied young stellar population. We based our sample on the survey conducted by \citet{fur08}. From this initial sample, we excluded CoKu\,Tau\,1, DG\,Tau\,B, HH\,30, IRAS\,04016+2610, IRAS\,04158+2805, IRAS\,04248+2612 and IRAS\,04302+2247, in which a nearly edge-on circumstellar disk strongly affects the system's SED \citep{bur96, pad99, gla08}. Indeed, in some extreme cases (HH\,30, IRAS\,04158+2805), an edge-on disk can make a Class\,II source erroneously appear like a Class\,I embedded object if one only focuses on the near- to mid-infrared spectral index.

For the 21 bona fide Class\,I sources in the sample, we adopted the broadband photometry presented in \citet{fur08} and complemented it with photometric measurements from the {\it AKARI} and {\it WISE} all-sky surveys whenever available. We also included {\it Herschel} photometry for a few individual sources from \cite{how13}, \cite{bul14}, and \cite{rib17} in cases where far-infrared observations were lacking. The resulting photometric database contains 455 individual flux density measurements for 21 objects at wavelength ranging 0.35\,$\mu$m to 1.3\,mm. 

The range of bolometric luminosities for this sample is very broad, from 0.25 to 28\,$L_\odot$ \citep{fur08}, such that distribution of the objects' flux densities at a given wavelength does not just reflect the diversity in the shape of Class\,I SEDs but also the range of their luminosities. To sidestep this problem, we transformed each object's SED to the unitless quantity $\lambda F_\lambda\,4\pi d^2\,/\,L_{bol}$ (where $d=140$\,pc), which effectively indicates the proportion of the object light output as a function of wavelength, i.e., the global shape of the SED. 

\begin{figure}
    \plottwo{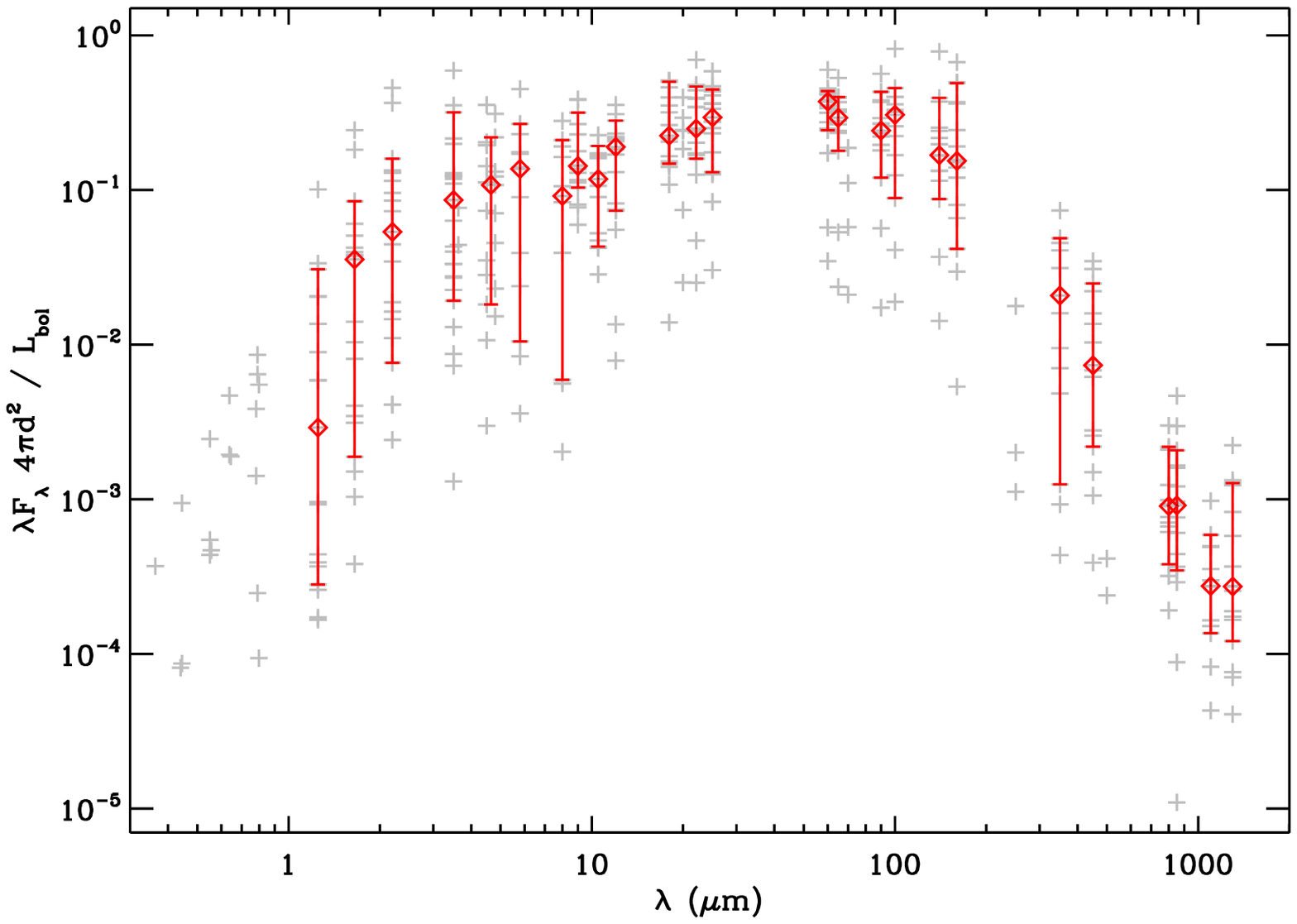}{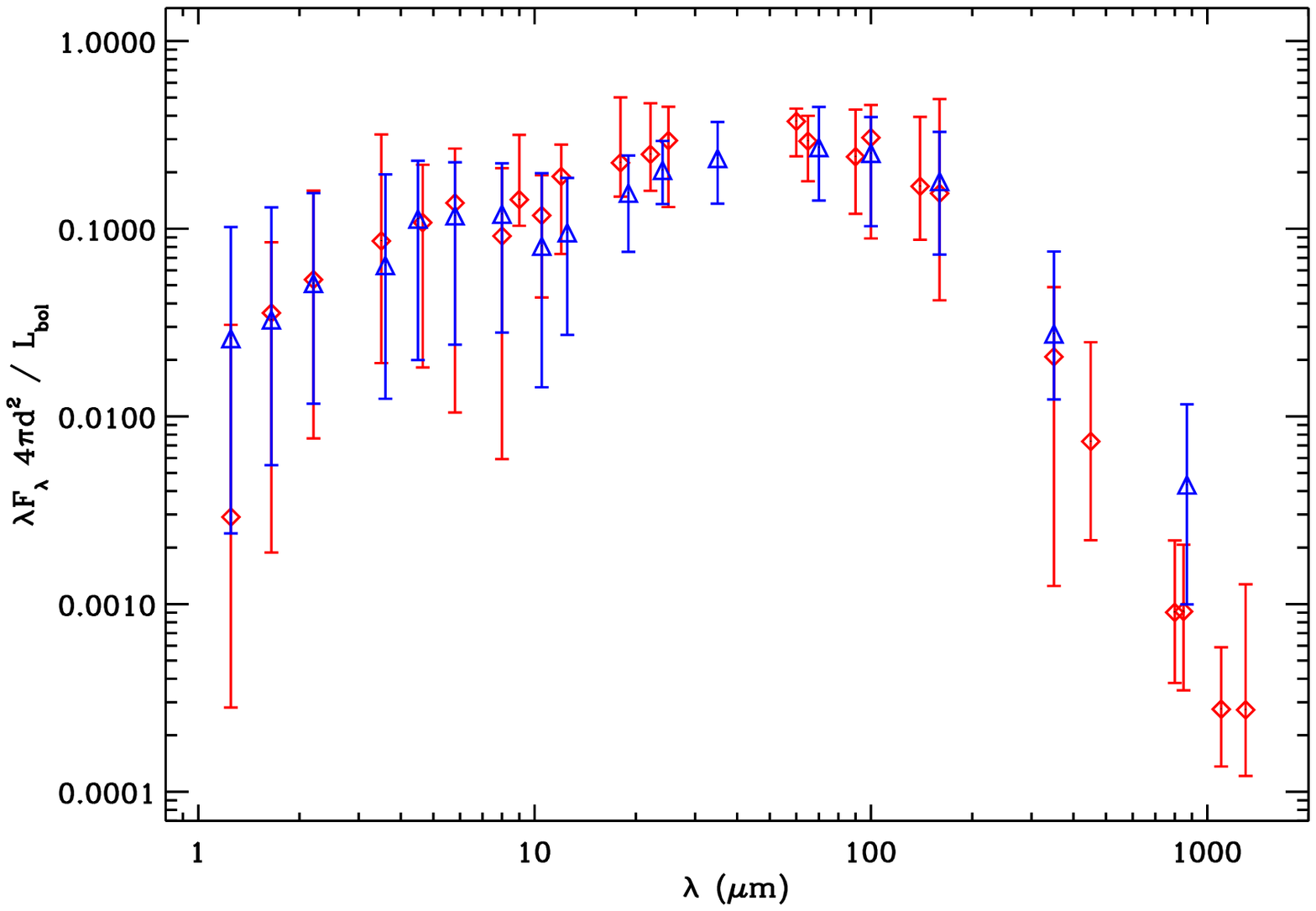}
    \caption{{\it Left:} Luminosity-normalized SEDs of Taurus Class\,I sources. Gray crosses represent individual photometric measurements whereas the red diamonds and errorbars indicate the median and 16-to-84 percentile range for wavelengths with observations of at least 10 objects (out of the sample of 21). {\it Right:} Luminosity-normalized SEDs of Orion Class\,I sources (blue triangles and errorbars indicate the median and 16-to-84 percentile range respectively). Red diamonds and errorbars reproduce the Taurus Class\,I template. \label{fig:sed_template}}
\end{figure}

We then grouped all luminosity-normalized measurements by observing wavelengths. At each wavelength for which more than 10 objects have a measured flux density (most notably, this limits our analysis to $\lambda \geq 1\,\mu$m), we then compute the median, 16 and 84 percentile. These estimates define our "template (Taurus) Class\,I SED" and associated uncertainties. Note that the latter rather represent the range of possible SED shapes, or intrinsic scatter within the population, rather than the actual measurement uncertainties. The template SED, along with all individual measurements, are shown in the left panel of Figure\,\ref{fig:sed_template} and listed in Table\,\ref{tab:sed_template}. The dispersion between objects is markedly larger at the longest and, especially, the shortest wavelengths whereas most objects are within a factor of $\approx$2 of the median SED in the mid- to far-infrared regime. This is a natural consequence of our normalization scheme and of the fact that this is the regime where most of the luminosity of a Class\,I is radiated away.

\begin{table}
        \centering
        \caption{Luminosity-normalized template SED for Class\,I sources in Taurus. The fifth column indicates the number of objects that contribute to the estimate of the median, 16 and 84 percentile of the distribution at each wavelength. Only wavelengths with $N_{\mathrm{obj}} \geq 10$ are listed. }
        \label{tab:sed_template}
        \begin{tabular}{cccccc}
          \hline
          \hline
        $\lambda$ & \multicolumn{3}{c}{$\lambda F_\lambda\,4\pi d^2\,/\,L_{\mathrm{bol}}$} & $N_{\mathrm{obj}}$ & Filter/Mission \\
         & 16 percentile & median & 84 percentile & & \\
          \hline
      1.25  & 0.000281 & 0.00291  & 0.0308   & 19 & 2MASS\,$J$ \\
      1.65  & 0.00188  & 0.0355   & 0.0846   & 20 & 2MASS\,$H$ \\
      2.20  & 0.00763  & 0.0536   & 0.160    & 21 & 2MASS\,$K$ \\
      3.6   & 0.0192   & 0.0861   & 0.318    & 19 & {\it Spitzer}/IRAC \\
      4.65  & 0.0182   & 0.108    & 0.219    & 17 & {\it Spitzer}/IRAC \\
      5.8   & 0.0105   & 0.137    & 0.267    & 13 & {\it Spitzer}/IRAC \\
      8.    & 0.00593  & 0.0915   & 0.210    & 11 & {\it Spitzer}/IRAC \\
      9.    & 0.104    & 0.143    & 0.317    & 16 & {\it AKARI} \\
      10.5  & 0.0430   & 0.118    & 0.193    & 15 & $N$ \\
      12.   & 0.0733   & 0.190    & 0.281    & 20 & {\it IRAS} \\
      18.   & 0.148    & 0.225    & 0.501    & 17 & {\it AKARI} \\
      22.1  & 0.159    & 0.249    & 0.466    & 18 & {\it WISE} \\
      25.   & 0.130    & 0.295    & 0.447    & 18 & {\it IRAS} \\
      60.   & 0.243    & 0.373    & 0.437    & 18 & {\it IRAS} \\
      65.   & 0.179    & 0.293    & 0.400    & 14 & {\it AKARI} \\
      90.   & 0.120    & 0.242    & 0.432    & 15 & {\it AKARI} \\
      100.  & 0.124    & 0.320    & 0.456    & 14 & {\it IRAS} \\
      140.  & 0.0874   & 0.168    & 0.394    & 15 & {\it AKARI} \\
      160.  & 0.0801   & 0.157    & 0.491    & 14 & {\it AKARI} \\
      350.  & 0.00705  & 0.0312   & 0.0488   & 10 &  \\
      450.  & 0.00219  & 0.00736  & 0.0249   & 17 &  \\
      800.  & 0.000380 & 0.000903 & 0.00218  & 14 &  \\
      850.  & 0.000347 & 0.000914 & 0.00207  & 17 &  \\
      1100. & 0.000136 & 0.000275 & 0.000589 & 13 &  \\
      1300. & 0.000121 & 0.000273 & 0.00127  & 17 &  \\
          \hline
        \end{tabular}   
\end{table}

While the Taurus Class\,I sources are the best studied sample available to build such a template, the sample is relatively small, and may not be representative of Class\,I sources in other star-forming regions. To address this issue, we followed the same procedure for the sample of 219 Orion Class\,I and Flat Spectrum sources from \cite{fur16}. As illustrated in the right panel of Figure\,\ref{fig:sed_template}, the general structure and ranges of the resulting Orion Class\,I template is consistent with the Taurus template. The only marginal difference between the two template is in the submillimeter regime, where the Orion template is higher than the Taurus one by a factor of a few. Most likely, this is a consequence of the enhanced crowding of the Orion region relative to Taurus, which is further compounded with confusion with cloud emission. These effect lead to a Malmquist bias whereby only the brightest sources can be detected in Orion. Indeed, in the submillimeter, only 20--30\% of all Orion sources are detected and included in the template, whereas in Taurus the corresponding fraction is roughly 70\%. Given the similarity of the two templates, we adopt for our study the more uniform and complete Taurus template.

\section{150 Variable sources in Orion}

We considered the 150 variable objects in Orion that had newly identified periods in \cite[][see their Table\,4]{mor11} to place the amplitude of variability of the Class\,I sources studied here in their broader context. Roughly half of these 150 objects were assigned the classification given to them in \cite{meg12}, where we considered "protostars" to be Class I (these are the same targets as studied in Section\,\ref{sec:ClassI}) and "disk-bearing" objects to be Class II. We then classified the remaining objects that were not assessed in \cite{meg12} as Class III.  Following the procedure described in Section \ref{sec:spot_fit}, the light curves from these 150 sources were fit to a single spotted model using the method of least-squares. 
The results from our light curve analysis, specifically the variability amplitude $\alpha$, are presented in Table \ref{tab:150lctab}. 

\begin{table}
\centering
\label{tab:orion_variables}

 \caption{The amplitude of variation pulled from the best fit to the object's light curve, listed with the object's name, class, and the reduced chi square value from the best-fitting light curve. Objects marked with $^{\dagger}$ are eclipsing binary candidates. Only the first 20 entries in the table are shown here; the full table is available electronically. \label{tab:150lctab}}
    \begin{tabular}{ccccc}
\hline
\hline
 Object Name & Class & P[d] & $\alpha$ & $\chi^2_{\mathrm{red}}$ \\
\hline
%Object Name               & Object Class & $\alpha$ & $\chi _{red}$\\ 
ISOY\:J053427.70-053155.4 & III & 3.37           & -0.05   & 2.63              \\
ISOY\:J053434.88-044243.5 & III & 9.36           & 0.05    & 3.03              \\
ISOY\:J053435.15-053210.4 & III&5.18            & -0.01   & 1.14              \\
ISOY\:J053435.98-045218.0 & II &    2.34        & 0.11    & 2.67              \\
ISOY\:J053439.11-053839.2 & III & 2.41           & 0.05    & 2.29              \\
ISOY\:J053439.76-052425.6 & III & 16.97          & 0.05    & 2.65              \\
ISOY\:J053441.18-054611.7 & II &    8.02       & 0.11    & 1.98              \\
ISOY\:J053441.71-052653.0 & II &    8.47       & 0.05    & 1.73              \\
ISOY\:J053441.87-055502.6 & II &    1.66       & -0.06   & 2.55              \\
ISOY\:J053442.74-052837.6 & II &    6.15        & -0.12   & 3.26              \\
ISOY\:J053445.00-045559.2 & II &    3.40       & -0.07   & 1.71              \\
ISOY\:J053447.64-054350.7 & III & 4.27          & 0.05    & 2.00              \\
ISOY\:J053447.74-052632.1 & II & 1.84           & 0.11    & 1.99              \\
ISOY\:J053447.85-044228.9 & III&    8.36        & -0.08   & 2.08              \\
ISOY\:J053449.57-052903.4 & II &    14.73       & -0.13   & 1.95              \\
ISOY\:J053450.66-050407.7 & III & 9.24          & 0.02    & 1.06              \\
ISOY\:J053450.72-045836.8 & II &    6.32       & 0.12    & 3.94              \\
ISOY\:J053452.22-045102.7 & III & 5.15          & -0.05   & 2.16              \\
ISOY\:J053452.49-044940.4 & III & 6.25            & -0.01   & 1.23              \\
ISOY\:J053452.60-051536.6 & III & 2.35            & 0.13    & 5.31              \\
\hline
\end{tabular}
\end{table}

%==========================================================

\label{lastpage}
\end{document}